# GYM: A Multiround Distributed Join Algorithm

Foto N. Afrati[1], Manas R. Joglekar[2], Christopher M. Re[2], Semih Salihoglu[3], and Jeffrey D. Ullman[2]

1   National Technical University of Athens, Athens, Greece
    afrati@gmail.com
2   Stanford University, California, USA
    manasrj@stanford.edu, chrismre@cs.stanford.edu, ullman@gmail.com
3   University of Waterloo, Ontario, Canada
    semih.salihoglu@uwaterloo.ca

**Abstract**

Multiround algorithms are now commonly used in distributed data processing systems, yet the extent to which algorithms can benefit from running more rounds is not well understood. This paper answers this question for several rounds for the problem of computing the equijoin of $n$ relations. Given any query $Q$ with width $w$, *intersection width* $iw$, input size IN, output size OUT, and a cluster of machines with $M = \Omega(\text{IN}^{\frac{1}{\epsilon}})$ memory available per machine, where $\epsilon > 1$ and $w \geq 1$ are constants, we show that:

1. $Q$ can be computed in $O(n)$ rounds with $O(n\frac{(\text{IN}^w+\text{OUT})^2}{M})$ communication cost with high probability.

2. $Q$ can be computed in $O(\log(n))$ rounds with $O(n\frac{(\text{IN}^{\max(w,3iw)}+\text{OUT})^2}{M})$ communication cost with high probability.

Intersection width is a new notion we introduce for queries and generalized hypertree decompositions (GHDs) of queries that captures how connected the adjacent components of the GHDs are.

We achieve our first result by introducing a distributed and generalized version of Yannakakis's algorithm, called GYM. GYM takes as input any GHD of $Q$ with width $w$ and depth $d$, and computes $Q$ in $O(d + \log(n))$ rounds and $O(n\frac{(\text{IN}^w+\text{OUT})^2}{M})$ communication cost. We achieve our second result by showing how to construct GHDs of $Q$ with width $\max(w, 3iw)$ and depth $O(\log(n))$. We describe another technique to construct GHDs with longer widths and lower depths, demonstrating other tradeoffs one can make between communication and the number of rounds.

## 1 Introduction

The problem of evaluating joins efficiently in distributed environments has gained importance since the advent of Google's MapReduce [10] and the emergence of a series of distributed systems with relational operators, such as Pig [25], Hive [29], SparkSQL [28], and Myria [19]. These systems are conceptually based on Valiant's *bulk synchronous parallel* (BSP) computational model [30]. Briefly, there are a set of machines that do not share any memory and are connected by a network. The computation is broken into a series of *rounds*. In each round, machines perform some local computation in parallel and communicate messages over the network. Costs of algorithms in these systems can be broken down to: (1) local computation of machines; (2) communication between the machines; and (3) the number of new rounds of computation that are started, which can have large overheads in some systems, e.g. due to reading input from disk or waiting for resources to be available in the cluster. In this paper, we focus on communication and the number of rounds, as for many data processing tasks, the computation cost is generally subsumed by the communication cost [20, 21].

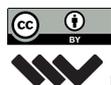

This paper studies the problem of evaluating an equijoin query $Q$ in multiple rounds of computation in a distributed cluster. We restrict ourselves to queries that are full, i.e. do not contain projections, but queries can contain self-joins. We let $n$ be the number of relations, IN the input size, OUT the output size of $Q$, and $M = o(\text{IN})$ the memory available per machine in the cluster. Memory sizes of the machines intuitively capture different parallelism levels: when memory sizes are smaller, we need a larger number of machines to evaluate the join, which increases parallelism. We assume throughout the paper that $M = \Omega(\text{IN}^{\frac{1}{\epsilon}})$ for some constant $\epsilon > 1$. For practical values of input and memory sizes, $\epsilon$ is a small constant. For instance, if IN is in terabytes, then even when $M$ is in megabytes, $\epsilon \approx 2$.

Our study of multiround join algorithms is motivated by two developments. First, it has been shown recently that there are prohibitively high lower bounds on the communication cost of any one-round algorithm for evaluating some join queries [3, 6]. For example, for the *chain query*, $C_n = R_1(A_0, A_1) \bowtie R_2(A_1, A_2) \bowtie \cdots \bowtie R_n(A_{n-1}, A_n)$, the lower bound on the communication cost of any one-round algorithm is $\geq (\frac{\text{IN}}{M})^{n/4}$. For example, if the input is one petabyte, i.e., IN=$10^{15}$, even when we have machines with ten gigabytes of memory, i.e., $M=10^{10}$, the communication cost of any one-round algorithm to evaluate the $C_{16}$ query is 100000 petabytes. Moreover, this lower bound holds even when the query output is known to be small, e.g., OUT = $O(\text{IN})$, and the input has no skew [6], implying that designing multiround algorithms is the only way to compute such joins more efficiently.

Second, the cost of running a new round of computation has decreased from several minutes in the early systems (e.g., Hadoop [5]) to milliseconds in some recent systems (e.g., Spark [32]), making it practical to run algorithms that consist of a large number of rounds. Although multiround algorithms are becoming commonplace, how much algorithms can benefit from running more rounds is not well understood. In this paper, we answer this question for equijoin queries.

We describe a multiround algorithm, called *GYM*, for **G**eneralized **Y**annakakis in **M**apReduce (Sections 4-5), which is a distributed and generalized version of Yannakakis's algorithm for acyclic queries [31]. The performance of GYM depends on two important structural properties of the input query: *depths* and *widths* of its *generalized hypertree decompositions* (GHDs). We then present two algorithms, *Log-GTA* (Section 6) and *C-GTA* (Section 7), for constructing GHDs of queries with different depths and widths, exposing a spectrum of tradeoffs one can make between the number of rounds and communication using GYM. In the remainder of this section, we give an overview of our results.

## 1.1 GYM: A Multiround Join Algorithm

The width of a query, i.e., the minimum width of any of its GHDs, characterizes its degree of cyclicity, where acyclic queries are equivalent to width-1 queries. The original serial algorithm of Yannakakis takes as input a width-1 GHD of an acyclic query. GYM generalizes Yannakakis's algorithm to take as input any GHD of any query $Q$ and evaluates $Q$ in a distributed fashion. In this paper, we focus on bounded width queries in this paper, i.e. those whose widths are a constant.

▶ Main Result 1. Given a width-$w$, depth-$d$ GHD of a query $Q$ over $n$ relations, GYM

| Query | Width | Min-Depth GHD | Intersection Width |
|---|---|---|---|
| $S_n : S(A_1, \ldots, A_{n-1}) \bowtie R_1(A_1, B_1) \bowtie \cdots \bowtie R_{n-1}(A_{n-1}, B_{n-1})$ | 1 | 1 | 1 |
| $C_n : R_1(A_0, A_1) \bowtie R_2(A_1, A_2) \bowtie \cdots \bowtie R_n(A_{n-1}, A_n)$ | 1 | $\Theta(n)$ | 1 |
| $TC_n : R_1(A_0, A_1) \bowtie R_2(A_0, A_2) \bowtie R_3(A_1, A_2) \bowtie$ $R_4(A_2, A_3) \bowtie R_5(A_2, A_4) \bowtie R_6(A_3, A_4) \bowtie$ ... $R_{n-2}(A_{\frac{2n}{3}-2}, A_{\frac{2n}{3}-1}) \bowtie R_{n-1}(A_{\frac{2n}{3}-2}, A_{\frac{2n}{3}}) \bowtie R_n(A_{\frac{2n}{3}-1}, A_{\frac{2n}{3}})$ | 2 | $\Theta(n)$ | 1 |

**Table 1** Example Queries $S_n$, $C_n$, and $TC_n$.

computes $Q$ in $O(\mathsf{d} + \log(n))$ rounds with $O(n\frac{(\text{IN}^{\mathsf{w}}+\text{OUT})^2}{M})$ communication cost with high probability.

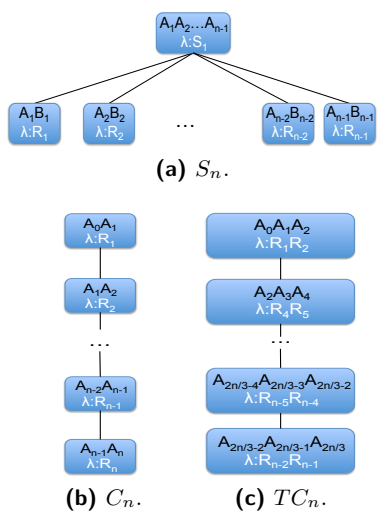

(a) $S_n$.

(b) $C_n$.  (c) $TC_n$.

**Figure 1** Example GHDs.

Since every width-w query over $n$ relations has a GHD of width w and depth at most $n$, an immediate corollary to our first main result is that every width-w query can be computed in $O(n)$ rounds and $O(n\frac{(\text{IN}^{\mathsf{w}}+\text{OUT})^2}{M})$ communication cost using GYM.

Table 1 lists three example queries and their widths w, minimum depths of their width-w GHDs, and intersection widths (explained momentarily). Figure 1 shows example GHDs of these queries. The labels on the vertices of the GHDs in Figure 1 are the $\lambda$ and $\chi$ values, following the notation in Section 3.

▶ **Example 1.** The star query $S_n$ is an acyclic query. As shown in Figure 1, $S_n$ has a depth-1 and width-1 GHD. Using this GHD, GYM executes $S_n$ in $O(\log(n))$ rounds with a communication cost of $O(n\frac{(\text{IN}+\text{OUT})^2}{M})$.

▶ **Example 2.** The chain query $C_n$ is also an acyclic query. Figure 1 shows an example width-1 GHD of $C_n$ with depth $n-1$. On this GHD, GYM executes $C_n$ in $O(n)$ rounds with a communication cost of $O(n\frac{(\text{IN}+\text{OUT})^2}{M})$.

We present GYM within the context of the MapReduce system because it is the earliest and one of the simplest modern large-scale data processing systems. However, GYM can easily run on any BSP system, so our results apply to other BSP systems as well. We also note that all of the results presented in this paper hold under any amount of skew in the input data. We discuss the improvements to our results when the inputs to queries are skew-free in Appendix C.

## 1.2 Log-GTA: Log-depth GHDs

For some width-w queries, any width-w GHD of the query has a depth of $\Theta(n)$. $C_n$ and the triangle-chain query, $TC_n$, shown in Table 1, are examples of such queries with widths 1 and 2, respectively. Therefore, on any width-w GHD of such queries, GYM executes $\Theta(n)$ rounds. Our second main result shows how to execute such queries by GYM in exponentially fewer number of rounds but with more communication cost by proving a combinatorial lemma about GHDs, which may be of independent interest to readers:

▶ Main Result 2. Given a width-w, intersection-width-iw, and depth-d GHD $D$ of $Q$, we can construct a GHD $D'$ of $Q$ of depth $O(\log(n))$ and width at most max(w, 3iw).

*Intersection width* is a new notion of GHDs we introduce, that captures how connected the adjacent components of a GHD are. We present an algorithm *Log-GTA*, for **Log**-*depth* **G**HD **T**ransformation **A***lgorithm*, to achieve our second main result. Using Log-GTA, we can tradeoff rounds and communication for queries with high depth GHDs as follows:

▶ Example 3. The $TC_n$ query has a width of 2 and intersection width of 1. Figure 1c shows an example width-2 GHD $D$ of $TC_n$ which has a depth of $\frac{n}{3}$-1. One option to evaluate $TC_n$ is to use $D$ directly. On $D$, GYM will execute $\Theta(n)$ rounds and have a communication cost of $O(n\frac{(\text{IN}^2+\text{OUT})^2}{M})$. Another option is to construct a new GHD $D'$ from $D$ by Log-GTA, which will have a depth of $O(\log(n))$ and width of 3. On $D'$, GYM will take $O(\log(n))$ rounds and have a communication cost of $O(n\frac{(\text{IN}^3+\text{OUT})^2}{M})$.

We end this section by discussing two interesting consequences of Log-GTA and GYM.

**Log-depth Decompositions.** GHDs [13] are one of several structural decomposition methods that are used to characterize the cyclicity of queries. Each decomposition method represents queries as a graph (if the input relations have arity at most 2) or a hypergraph and has a notion of "width" to measure the cyclicity of queries. Examples include *query decompositions* [8], *tree decompositions* (TDs) [27], and *hypertree decompositions* (HDs) [18]. Two previous results by Bodlaender [7] and Akatov [4] have proved the existence of log-depth TDs of hypergraphs with thrice their *treewidths* and HDs of hypergraphs with thrice their *hypertreewidths*, respectively. Our second main result proves that a similar and stronger property also holds for GHDs of hypergraphs. Interestingly, neither of these results (including ours) imply each other. However, we show in Appendix G.2 that Log-GTA also recovers Bodlaender's result. That is, Log-GTA also transforms a given TD into log-depth one with thrice its treewidth. In addition, we show that a modification of Log-GTA recovers both Akatov's and Bodlaender's results and a weaker version of our second main result. We also show that using similar definitions of intersection widths for TDs and HDs, we can improve both Bodlaender's and Akatov's results.

**Parallel Complexity of Bounded-width Queries.** Database researchers have often thought of Yannakakis's algorithm as having a sequential nature, executing for $\Theta(n)$ steps in the PRAM model. In the PRAM literature [11, 16, 17], acyclic queries have been described as being polynomial-time sequentially solvable by Yannakakis's algorithm, but highly parallelizable by the *ACQ* algorithm [15], where parallelizability refers to being in the complexity class NC. By constructing log-depth GHDs of queries and simulating GYM in the PRAM model, we show that unlike previously thought, with simple modifications Yannakakis's algorithm can run in logarithmic rounds, implying that bounded-width queries are in the complexity class NC, recovering a result that was proven using the ACQ algorithm [15]. We note that BSP models can also simulate PRAM and a comparison of GYM against a distributed simulation of ACQ, which we call *ACQ-MR*, is given in Appendix B. As we show in Appendix B, GYM can match ACQ-MR's performance on every query using an appropriate GHD for the query and on some queries GYM strictly outperforms ACQ-MR.

## 2 Related Work

We provide a brief overview of related work here. A comprehensive coverage of related work is in the full version of this paper [1].

**Shares.** Shares [3] is the optimal one-round join algorithm. References [2] and [6] have shown that for every query $Q$, and value of $M$, and skew level, the Shares algorithm can be configured to incur the lowest possible communication cost among one-round algorithms that send at most $M$ input tuples to each machine. However, as we discussed in Section 1, for some queries, these costs can be prohibitively expensive.

**Other Distributed Join Algorithms.** Reference [6] studies multiround distributed join algorithms in the *Massively Parallel Computing* (MPC) model. Reference [6] proves lower bounds on the number of rounds required to compute queries when $M = \frac{\text{IN}}{p^{1-\epsilon'}}$, where $p$ is the number of machines, and $\epsilon'$ is a constant $\in [0, 1)$ called the *space exponent*. They show that when evaluating a query whose GHDs are of depth d on an arbitrary database, any algorithm with limited memory, where the limitation is defined as space exponent being a constant, will have to run $O(\log(\mathsf{d}))$ rounds. The authors show that running the Shares algorithm iteratively on sets of the input relations matches these lower bounds on a limited set of inputs, called *matching databases*, which represent skew-free inputs. On arbitrary databases, their iterative Shares algorithm can produce intermediate data of size $\text{IN}^{\Theta(n)}$ for any query irrespective of its width. By our second result, GYM evaluates these queries on any database instance in $O(\log(n))$ rounds. So when d i constant but $n$ is unbounded, GYM runs $O(\log(n))$ rounds whereas their lower bound is $O(1)$. However, GYM keeps intermediate relation sizes bounded by $\text{IN}^{\max(w, 3\mathsf{iw})} + \text{OUT}$. On matching databases, their algorithms matches these lower bounds exactly. Similarly, our GYM algorithm matches these lower bounds exactly using the optimizations we outline in Appendix C.

Reference [23] describes worst case optimal constant-round join algorithms for several classes of conjunctive queries when the frequencies of each value in the attributes of the relations are known. The authors relate the communication cost of algorithms on a query to a structural property of the query called *edge quasi-packing number*, which can be smaller than the width of the query. In contrast, we do not assume any prior knowledge of frequencies.

**Generalized Hypertree Decompositions.** Structural decomposition methods, such as GHDs [14], query decompositions (QDs) [8], tree decompositions (TDs) [27], and hypertree decompositions (HDs) [18], are mathematical tools to characterize the difficulty of computational problems that can be represented as graphs or hypergraphs, such as joins or constraint satisfaction problems. GYM can use methods other than GHDs, such as QDs, and HDs, but we use GHDs because the widths of GHDs are known to be smaller than HDs and QDs, giving us stronger results in terms of communication cost.

## 3 Preliminaries

We review GHDs, describe our model, and specify the assumptions we make in this paper.

### 3.1 Generalized Hypertree Decompositions

A **hypergraph** is a pair $H = (V(H), E(H))$, consisting of a nonempty set $V(H)$ of vertices, and a set $E(H)$ of subsets of $V(H)$, the hyperedges of $H$. Natural join queries can be expressed as hypergraphs, where we have a vertex for each attribute of the query, and a hyperedge for each relation.

▶ **Example 4.** Consider the query Q:

$$R_1(A, B, C) \bowtie R_2(B, F) \bowtie R_3(B, C, D) \bowtie \\ R_4(C, D, E) \bowtie R_5(D, E, G)$$

The hypergraph of $Q$ is shown in Figure 2a.

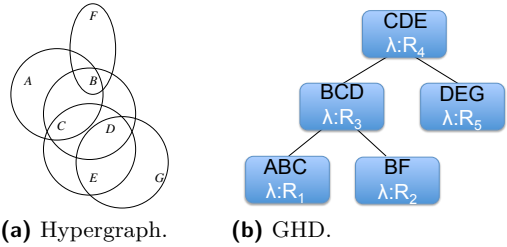

(a) Hypergraph.  (b) GHD.

**Figure 2** Hypergraph and GHD of Example 4.

Let $H$ be a hypergraph. A ***generalized hypertree decomposition (GHD)*** of $H$ is a triple $D = (T, \chi, \lambda)$, where:
- $T(V(T), E(T))$ is a tree;
- $\chi : V(T) \to 2^{V(H)}$ is a function associating a set of vertices $\chi(t) \subseteq V(H)$ to each vertex $t$ of $T$;
- $\lambda : V(T) \to 2^{E(H)}$ is a function associating a set of hyperedges to each vertex $t$ of $T$;

such that the following properties hold:

1. For each $e \in E(H)$, there is a vertex $t \in V(T)$ such that $e \subseteq \chi(t)$.
2. For each $v \in V(H)$, the set $\{t \in V(T) | v \in \chi(t)\}$ is connected in $T$.
3. For every $t \in V(T)$, $\chi(t) \subseteq \bigcup \lambda(t)$, i.e., hyperedges of $\lambda(t)$ must "cover" the vertices of $\chi(t)$.

For any $t \in V(T)$, we refer to $\chi(t)$ as the attributes of $t$ and $\lambda(t)$ as the relations on $t$. A GHD of a join query $Q$ is defined to be a GHD on the hypergraph of $Q$.

▶ **Example 5.** Figure 2b shows a GHD of the query from Example 4. In the figure, the attribute values on top of each vertex $t$ are the $\chi$ assignments for $t$ and the $\lambda$ assignments are explicitly shown.

We next define several properties of GHDs and hypergraphs:
- The ***depth*** of a GHD $D = (T, \chi, \lambda)$ is the depth of the tree $T$.
- The ***width*** of a GHD $D$ is $\max_{t \in V(T)}\{|\lambda(t)|\}$, i.e., the maximum number of relations assigned to any vertex $t$.
- The ***generalized hypertree width***, or ***width*** for short, of a hypergraph $H$ is the minimum width of all GHDs of $H$.

The width of a query captures its degree of cyclicity. In general, the larger the width of a query, the more "cyclic" it is. Acyclic queries are exactly the queries with width 1 [8]. We next define a new notion called intersection width.
- The ***intersection width*** of a GHD $D = (T, \chi, \lambda)$ is defined as follows: For any adjacent vertices $t, t' \in V(T)$, let $\mathsf{iw}(t, t')$ denote the size of the smallest set $S \subseteq E(H)$ such that $\chi(t) \cap \chi(t') \subseteq \bigcup_{s \in S} s$. In other words, $\mathsf{iw}(t, t')$ is the size of the smallest set of relations whose attributes cover the common attributes between $t$ and $t'$. The intersection width iw of a GHD is the maximum $\mathsf{iw}(t, t')$ over all adjacent $t, t' \in V(T)$.

Notice that the intersection width of $D$ is never larger than the width of $D$, because $\forall\, t, t' \in V(T) : \mathsf{iw}(t, t') \leq |\lambda(t)|$, since by the 3rd property of GHDs $\lambda(t)$ is one (possibly not the smallest) set of relations that covers the attributes of $t$, and therefore any common attribute that $t$ shares with its neighbors. The intersection width of a GHD can be strictly smaller than the width, as the next example shows.

▶ **Example 6.** Consider the $TC_n$ example from Table 1. $TC_n$ is a width-2 query. The hypergraph of $TC_n$, shown in Figure 3, visually is a chain of triangles, where any two consecutive triangles are connected by a single attribute. Figure 1c shows a width-2 GHD of $TC_n$, where each node covers one of the triangles in the same order they appear in the hypergraph. The intersection width of this GHD is 1, as the common attribute between each triangle can be covered by one relation (e.g., $A_2$, which is the common attribute between the first two triangles, can be covered by $R_2$).

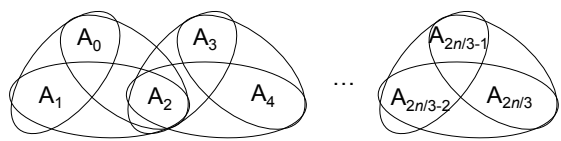

**Figure 3** Hypergraph of $TC_n$.

In the rest of this paper we restrict ourselves, for simplicity of presentation, to queries whose hypergraphs are connected. All of our results generalize to queries with disconnected hypergraphs. A GHD $D(T, \chi, \lambda)$ of a hypergraph $H$ is called *complete* if each hyperedge $e \in E(H)$ occurs in $\lambda(t)$ of some vertex $t \in V(T)$. That is, each relation is assigned to the $\lambda$-label of some vertex $t \in V(T)$. We assume throughout the paper that the GHDs we use are rooted, i.e., one of the vertices (arbitrarily) in $T$ is picked as a root. This ensures that there is a well defined notion of height of vertices and parent-child relationships between the vertices in $T$. We end this section by stating a lemma about complete GHDs of queries:

▶ **Lemma 7.** *If a query $Q$ has a width-*w*, intersection width-*iw *GHD $D = (T, \chi, \lambda)$ of depth* d*, then $Q$ has a complete GHD $D' = (T', \chi', \lambda')$ with depth $\leq d+1$, width* w*, intersection width* iw*, and $|V(T')| \leq 4n$.*

We prove this lemma in Appendix D.1. Using this lemma, we will assume w.l.o.g. that the GHDs of queries that we use in our algorithms are complete and have $O(n)$ size.

## 3.2 MapReduce and Cost Model

Our MapReduce (MR) model is equivalent to the MR model in reference [26] except we use tuples instead of bits as our cost unit. In the tuple-based MR model, the unit of memory and communication cost is a base or intermediate tuple consisting of any set of attributes in $Q$. There is a set of distributed machines on a networked file system, each with memory $M = o(\text{IN})$.

**Map Stage**: Each machine, referred to as a *mapper*, reads a set of base or intermediate tuples over any set of attributes in the query from the networked file system. Mappers can send tuples to one or more machines, called *reducers*[1], deterministically or by hashing them on any set of their attributes. We assume that mappers have access to the same random bits and families of universal hash functions.[2] Suppose there are $k$ reducers. Using an appropriate family of hash functions, mappers can hash tuples of an input or intermediate relation $R$, using any subset of the tuples' attributes, to one of the $k$ reducers. Therefore two tuples with the same attributes that were used in the hashing will go to the same reducer. The total number of tuples received by a reducer from all mappers should not exceed memory size $M$. Otherwise the computation aborts. We note that we use randomization for load-balancing only. Specifically, GYM might send a machine more than $M$ tuples, exceeding the memory capacity of the machine, resulting in the computation to abort. However, this will happen with exponentially small probability.

---

[1] The machines we refer to as reducers are equivalent to *reduce keys* in the original description of MR [10]. These are separate groups of data on which the *reduce()* function is executed in the original system.
[2] Access to these random bits do not require any synchronization. In practice this would achieved by using a pseudorandom number generator with the same seed.

**Reduce Stage:** Each reducer locally performs any computation on the $\leq M$ tuples it receives, produces a set of output tuples, and streams the output tuples to the network file system. The local computation at a reducer cannot exceed memory size $M$, but the output of a reducer can exceed $M$ as it is streamed to the file system.

The *communication cost* of each round is defined as the total number of tuples sent from all mappers to reducers plus the number of output tuples produced by the reducers. We measure the complexity of our algorithms in terms of the total communication cost and the number of rounds. In Appendix A, we compare our model to existing models of modern distributed BSP systems.

### 3.3 Assumptions

We next specify three assumptions we make throughout the paper.

1. As in many MapReduce and distributed BSP models [6, 12, 22, 24, 26], we constrain $M$ to be $o(\text{IN})$. This ensures that machines cannot store the entire input. Otherwise we can send the entire input to a single machine and incrementally evaluate the join using any binary join plan, without exceeding memory $O(M)$, in a single round and without any communication (except to write the output to the networked file system).
2. We assume $M = \Theta(\text{IN}^{\frac{1}{\epsilon}})$ for some constant $\epsilon > 1$.
3. We assume queries have constant widths, i.e., the term w is a constant ($O(1)$).

As we discuss in Section 5, the complexities of our algorithms become slightly worse when we drop assumptions (2) and (3).

### 3.4 Basic Relational Operations in MR

We next state four lemmas characterizing the costs of joins, duplicate elimination, semijoins, and intersections in our model.

▶ **Lemma 8.** *Any $z$ relations $R_1, \ldots, R_z$ can be joined in $1$ round with $O(\frac{z^z(\Sigma_i|R_i|)^z}{M^{z-1}} + |R_1 \bowtie \cdots \bowtie R_z|)$ communication. When $z$ is a constant, the join can be performed in $O(\frac{(\Sigma_i|R_i|)^z}{M^{z-1}} + |R_1 \bowtie \cdots \bowtie R_z|)$ communication.*

**Proof.** We perform the join as follows. We divide each $R_i$ in $g_{r_i} = \frac{z|R_i|}{M}$ disjoint groups of size $\frac{M}{z}$ each. Then we use a total of $g_{r_1} \times \cdots \times g_{r_z}$ reducers and map a distinct set of $z$ groups, one from each relation, to each reducer. Each reducer joins its $z$ groups locally. Thus we use $\frac{z^z|R_1|\ldots|R_z|}{M^z}$ reducers and each reducer gets an input equal to $M$, and the total output size is $|R_1 \bowtie \cdots \bowtie R_z|$. Therefore, the total communication cost is $O(\frac{z^z(\Sigma_i|R_i|)^z}{M^{z-1}} + |R_1 \bowtie \cdots \bowtie R_z|)$. ◀

We note that Lemma 8 holds even if the given relations contain self-joins.

▶ **Lemma 9.** *Let $S$ be a multiset such that each tuple $t \in S$ has at most $k$ duplicates. We can remove the duplicates in $S$ w.h.p. in $O(\log_M(k))$ rounds and $O(\log_M(k)|S|)$ communication.*

**Proof.** We cannot send the duplicates of each tuple to a separate reducer because $k$ might be greater than $M$, exceeding the memory of machines. Let $h$ be a hash function mapping the tuples of $S$ into $|S|^2$ buckets randomly, so w.h.p. each bucket $h(i)$ gets $O(1)$ unique tuples. In the first round of duplicate elimination, we use $|S|^2 k^2$ reducers indexed with two numbers, $(1,1), \ldots, (1,k^2), \ldots, (|S|^2, 1), \ldots, (|S|^2, k^2)$, and each tuple $t$ is mapped to the reducer with the first index $h(t)$ and a uniformly random second index. Therefore, w.h.p., each reducer gets $O(1)$ tuples. Reducers do not perform any computation on their tuples. Note that every duplicate of tuple $t$ is mapped to a reducer with the first index $h(t)$. In addition, the number of non-duplicate tuples across all of the reducers with the same first index is $O(1)$, since $h$ maps $O(1)$ unique $S$ tuples to each $h(i)$. In the second round, for each set of reducers with the same first index we do the following in parallel: We group the reducers into groups of $\sqrt{M}$, map their tuples to the same reducer, and eliminate the duplicates across them.[3] This reduces the number of reducers that can contain duplicates to $\frac{k^2}{\sqrt{M}}$. We then repeat this procedure in parallel for each group of reducers with the same first index until all duplicates within each group are eliminated. In each round, each reducer gets $O(\sqrt{M})$ tuples and outputs $O(1)$ tuples. This computation takes $O(\log_{\sqrt{M}}(k))$ rounds. Since the communication in each round is $|S|$, this computation takes $O(\log_{\sqrt{M}}(k)|S|)$ communication.[4]  ◂

▶ **Lemma 10.** *Let $B(X, M) = \frac{X^2}{M}$. Given two relations $R$ and $S$, the semijoin $S \ltimes R$ can be computed w.h.p. in $O(\log_M(|R|))$ rounds with $O(\log_M(|R|)B(|R| + |S|, M))$ communication. When $M = \Omega((|R|)^{1/\epsilon})$, for some $\epsilon > 1$, the semijoin can be performed in $O(1)$ rounds and $O(B(|R| + |S|, M))$ communication.*

**Proof.** The first round of the semijoin $S \ltimes R$ is similar to the join. Let $g_r = \frac{2|R|}{M}$ and $g_s = \frac{2|S|}{M}$ be disjoint groups of size $\frac{M}{2}$. Each of the $g_r g_s$ reducers locally computes the semijoin of one S group and one R group it receives. Because each tuple of $S$ is sent to $g_r$ different reducers, there may be up to $g_r$ duplicates of each tuple. So the size of the multiset $S'$ with duplicates of $S$ is $g_r|S|$. Using Lemma 9, we can eliminate these tuples w.h.p. in $O(\log_M(g_r))$ rounds and $O(\log_M(g_r)g_r|S|) = O(\frac{\log_M(|R|)|R||S|}{M})$ communication. Together with the costs of the join operation, which takes 1 round and $O(\frac{(|R|+|S|)^2}{M})$ communication, we conclude that the semijoin can be performed w.h.p. in $O(\log_M(|R|))$ rounds and $O(\log_M(|R|)B(|R| + |S|, M))$ communication cost. When $M = \Omega((|R|)^{1/\epsilon})$, the semijoin can be computed in $O(1)$ rounds and $O(B(|R| + |S|, M))$ communication.  ◂

▶ **Lemma 11.** *Two relations $R$ and $S$ can be intersected w.h.p in 1 round with $O(|R| + |S|)$ communication.*

**Proof.** Suppose w.l.o.g. that $|R| > |S|$. We simply hash each tuple of $R$ and $S$ using all of their attributes into $|R|^2$ machines, so w.h.p. each machine gets $O(1)$ tuples from both $R$ and $S$ and performs a local intersection, without exceeding $M$, and writes the at most $|R| + |S|$ output.  ◂

---

[3] We can also group them into $\frac{M}{c}$ for a constant $c$ that is larger than the ($O(1)$) tuples any reducer has.
[4] As is standard, we use the term w.h.p. in this paper to refer to probabilities that are exponentially small in the input size IN. The probabilities mentioned in Lemma 9 are exponentially small in the size of $S$, which as we will see, in Yannakakis's algorithm can be smaller than IN. However when an input relation $S$ on which we perform duplicate elimination is small, we can make this probability exponentially small in IN by simply increasing the number of our buckets to $\text{IN}^2$.

## 4 Distributed Yannakakis

We first review the serial version of Yannakakis's algorithm for acyclic queries in Section 4.1. In Section 4.2, we show how to run Yannakakis's algorithm in a distributed setting in $O(n)$ rounds and $O(n\text{B}(\text{IN}+\text{OUT}, M))$ communication cost. In Section 4.3, we reduce the number of rounds to $O(\mathsf{d} + \log(n))$ rounds without affecting the communication cost.

### 4.1 Serial Yannakakis Algorithm

The serial version of Yannakakis's algorithm takes as input an acyclic query $Q = R_1 \bowtie R_2 \bowtie \cdots \bowtie R_n$, and constructs a width-1 GHD $D = (T, \chi, \lambda)$ of $Q$. Since $D$ is a GHD with width 1, each vertex of $D$ is assigned exactly one relation $R_i$. We will refer to relations that are assigned to leaf (non-leaf) vertices in $T$ as *leaf (non-leaf) relations*. Yannakakis's algorithm first eliminates all tuples that will not contribute to the final output by a series of semijoin operations. The overall algorithm consists of two phases: (1) a **semijoin phase**; and (2) a **join phase**.

**Semijoin Phase:** The semijoin phase operates recursively as follows.

- BASIS: If $T$ is a single node, do nothing.
- INDUCTION: If $T$ has more than one node, pick a leaf $t$ that is assigned relation $R$, and let $S$ be the relation assigned to $t$'s parent.
  1. Replace $S$ by the semijoin of $S$ with $R$, $S \ltimes R = S \bowtie \pi_{R \cap S}(R)$.
  2. Recursively process $T \setminus R$.
  3. Compute the final value of $R$ by computing its semijoin with the value of $S$ that results from step (2); that is, $R := R \ltimes S$.

The executions of step (1) in this recursive algorithm form the *upward* sub-phase, and the executions of step (3) form the *downward* sub-phase. In total, this version of the algorithm performs $2(n-1)$ semijoin operations.

**Join Phase:** The algorithm performs a series of $(n-1)$ joins, in any bottom-up order on $T$.

An important property of Yannakakis's algorithm is that the semijoin phase removes all of the "dangling" tuples in the input, i.e., those that will not contribute to the final output. This guarantees that the sizes of all intermediate tables during the join phase are no larger than the final output size OUT [31].

### 4.2 DYM-n

If we simply execute each semijoin and join operation of Yannakakis's algorithm by the algorithms in Lemmas 8 and 10, we get a distributed algorithm which we refer to as ***DYM-n***:

▶ **Theorem 12.** *DYM-n computes every acyclic query $Q$ in $O(n)$ rounds and in $O(n\text{B}(\text{IN}+\text{OUT}, M))$ communication cost.*

**Proof.** The algorithm executes a total of $2(n-1)$ pairwise semijoins and $n-1$ joins, in a total of $O(n)$ rounds. The largest input to any semijoin operation is the largest relation size, which is at most IN. By Lemma 10, the communication cost of the semijoin phase is $O(n\text{B}(\text{IN}, M))$. In each round of the join phase, the input and outputs are at most the final output size OUT. By Lemma 8, the cost of each round is $O(\frac{\text{OUT}^2}{M} + \text{OUT})$. Therefore, the total cost of both phases is $O(n(\text{B}(\text{IN}, M) + \text{B}(\text{OUT}, M) + \text{OUT}))$, which is $O(n\text{B}(\text{IN}+\text{OUT}, M))$ when $M = O(\text{IN})$, as we assume in this paper (recall Section 3.3). ◀

## 4.3 DYM-d

*DYM-d* parallelizes Yannakakis's algorithm further by executing multiple semijoins and joins in parallel, reducing the number of rounds to $O(\mathsf{d} + \log(n))$, where $\mathsf{d}$ is the depth of the GHD $D(T, \chi, \lambda)$, without asymptotically affecting DYM-n's communication cost.

**Upward Semijoin Sub-phase in $O(\mathsf{d} + \log(n))$ Rounds:** During the upward semijoin sub-phase, the algorithm from Section 4.1 picks one leaf $t$ that is assigned relation $R$ and processes $R$ by replacing it's parent $S$ with $S \ltimes R$ in $O(1)$ rounds. Instead we can pick and process all leaves in parallel. Consider the set $L$ of leaves of $T$. Let $L_1$ be the set of leaves that have no siblings, and let $L_2$ be the remaining leaves. We will replace step (1) of the algorithm from Section 4.1 with two steps, which will be performed in parallel.

1.1. For each $R$ in $L_1$ in parallel, replace $R$'s parent $S$ with $S \ltimes R$.
1.2. Divide the leaves in $L_2$ into disjoint pairs of siblings, and up to one triple of siblings per parent, if there is an odd number of siblings with the same parent. Then in parallel perform the following computation. Suppose $R_1$ and $R_2$ form such a pair with parent $S$. Replace $R_1$ with $(S \ltimes R_1) \cap (S \ltimes R_2)$ and remove $R_2$. If there is a triple $R_1, R_2, R_3$, replace $R_1$ with $(S \ltimes R_1) \cap (S \ltimes R_2) \cap (S \ltimes R_3)$ (using two pairwise intersections) and remove $R_2$ and $R_3$.

▶ **Lemma 13.** *The above procedure runs in $O(\mathsf{d} + \log(n))$ rounds.*

The proof of this lemma is provided in Appendix D.2. Since we perform $O(n)$ intersection or semijoin operations in total and all of the initial and intermediate relations involved have size at most IN, by Lemmas 10 and 11, the total communication cost of the upward semijoin sub-phase is $O(n\mathrm{B}(\mathrm{IN}, M))$.

**Downward Semijoin Sub-phase in $O(\mathsf{d})$ Rounds:** Note that in the downward semijoin sub-phase, the semijoins of the children relations with the same parent are independent and can be done in parallel in $O(1)$ rounds. Thus we can perform the downward sub-phase in $O(\mathsf{d})$ rounds and in $O(n\mathrm{B}(\mathrm{IN}, M))$ communication.

**Join Phase in $O(\mathsf{d} + \log(n))$ Rounds:** The join phase is similar to the upward semijoin sub-phase. The only difference is, we compute $S \bowtie R$ instead of $S \ltimes R$ for $R \in L_1$, and $(R_1 \bowtie S) \bowtie (R_2 \bowtie S)$ for pair $R_1, R_2 \in L_2$. The total number of rounds required is again $O(\mathsf{d} + \log(n))$. The total communication cost of each pairwise join is $O(n\mathrm{B}(\mathrm{OUT}, M))$, since the intermediate relations being joined are at most as large as OUT. Therefore, both the semijoin and join phases can be performed in $O(\mathsf{d} + \log(n))$ rounds with a total communication cost of $O(n\mathrm{B}(\mathrm{IN} + \mathrm{OUT}, M))$, justifying the following theorem:

▶ **Theorem 14.** *DYM-d evaluates an acyclic query $Q$ in $O(\mathsf{d} + \log(n))$ rounds and $O(n\mathrm{B}(\mathrm{IN} + \mathrm{OUT}, M))$ communication cost, where $\mathsf{d}$ is the depth of a width-1 GHD $D(T, \chi, \lambda)$ of $Q$.*

## 5 GYM

Our *GYM* algorithm generalizes DYM-d from acyclic queries to any query. Consider a width-w, depth-d GHD $D(T, \chi, \lambda)$ of a query $Q$. By Lemma 7, we assume w.l.o.g. that $D$ is complete. Consider "materializing" each $v \in V(T)$ by computing $IDB_v = \bowtie_{R_i \in \lambda(v)} R_i$. Now, consider the query $Q' = \bowtie_{v \in V(T)} IDB_v$. Note that $Q'$ has the exact same output as $Q$. This is because $Q'$ is also the join of all $R_i$, where some $R_i$ might (unnecessarily) be joined multiple times if they are assigned to multiple vertices. However, observe that $Q'$ is now an acyclic query and $D$ is now a width-1 GHD for $Q'$. Therefore we can directly run DYM-d to compute $Q'$.

▶ **Theorem 15** (**First Main Result**). *Given a width-w, depth-d GHD $D(T, \chi, \lambda)$ of a query $Q$ over $n$ relations, GYM executes $Q$ in $O(\mathsf{d} + \log(n))$ rounds and $O(n\mathrm{B}(\mathrm{IN}^\mathsf{w} + \mathrm{OUT}, M))$ communication cost.*

**Proof.** For the materialization stage, for each vertex $v$ of $D$, joining the w relations inside $\lambda(v)$ takes 1 round and $O(\frac{\mathrm{IN}^\mathsf{w}}{M^{w-1}} + |IDB_v|)$ communication cost by Lemma 8. In the worst case when the relations constitute a Cartesian product, $|IDB_v|$ is $\mathrm{IN}^w$, so evaluating $IDB_v$ takes $O(\mathrm{IN}^\mathsf{w})$ cost. By Lemma 7 there are at most $4n$ vertices in $V(T)$, so the materialization stage takes $O(n\mathrm{IN}^\mathsf{w})$ communication cost. Since the size of each $IDB_v$ is at most $\mathrm{IN}^\mathsf{w}$, executing DYM-d on the $IDB_v$'s takes $O(\mathsf{d} + \log(n))$ rounds and $O(n\mathrm{B}(\mathrm{IN}^\mathsf{w} + \mathrm{OUT}, M))$ communication, which dominates the cost of materialization phase, completing the proof. ◀

In Appendix E, we present an example execution of GYM on a query. We note that if we drop our assumptions that $M = \Omega(\mathrm{IN}^{\frac{1}{\epsilon}})$ and w is a constant, the number of rounds that GYM takes on a width-w, depth-d GHD increases by a factor of $\log_M(\mathrm{IN}^\mathsf{w}) = \mathsf{w} \log_M(\mathrm{IN})$. This is because the semijoin operations on $O(\mathrm{IN}^\mathsf{w})$ size inputs will execute $O(w \log_M(\mathrm{IN}))$ rounds instead of $O(1)$. Similarly, the communication cost of GYM will increase by at most a factor of $\max\{\mathsf{w} \log_M(\mathrm{IN}), \mathsf{w}^\mathsf{w}\}$. The $\mathsf{w} \log_M(\mathrm{IN})$ and $\mathsf{w}^\mathsf{w}$ factors are due to the communication cost increases in the semijoin (Lemma 10) and join operations (Lemma 8), respectively. .

## 6 Log-GTA

We now describe our ***Log-GTA*** algorithm (for **Log**-depth **G**HD **T**ransformation **A**lgorithm) which takes as input a hypergraph $H$ of a query $Q$, and its GHD $D(T, \chi, \lambda)$ with width w and intersection width iw, and constructs a GHD $D^*$ with depth $O(\log(|V(T)|))$ and width $\leq \max(\mathsf{w}, 3\mathsf{iw})$. For simplicity, we will refer to all GHDs during Log-GTA's transformation as $D'(T', \chi', \lambda')$, i.e., $D' = D$ in the beginning and $D' = D^*$ at the end. By running GYM on $D^*$, we can execute $Q$ in $O(\log(n))$ rounds with $O(n\mathrm{B}(\mathrm{IN}^{\max(\mathsf{w},3\mathsf{iw})} + \mathrm{OUT}, M))$ communication.

### 6.1 Extending to D′

Log-GTA associates two new labels with the vertices of $T'$:

1. **Active/Inactive**: An 'active' vertex is one that will be modified in later iterations of Log-GTA. Log-GTA starts with all vertices active, and inactivates vertices iteratively until all of them are inactive. At any point, we refer to the subtree of $T'$ consisting of only active vertices as ***active(T′)***. We prove that active($T'$) is indeed a tree in Lemma 17.

2. **Height**: The height of a vertex is its minimum distance from a leaf of the tree. The height of each vertex $v$ is assigned when $v$ is first inactivated, and remains unchanged thereafter.

In addition, Log-GTA associates a label with each "active" edge $(u, v) \in E(active(T'))$:

- **Common-cover(u, v) (cc(u, v))**: Is a set $S \subseteq E(H)$ such that $(\chi(u) \cap \chi(v)) \subseteq \bigcup_{s \in S} s$. In query terms, cc($u, v$) is a set of relations whose attributes cover the common attributes between $u$ and $v$. In the original $D(T, \chi, \lambda)$, for each $(u, v)$, we set cc($u, v$) to any covering subset of size at most iw. Recall from Section 3 that by definition of iw, such a subset must exist.

**Unique-c-gc vertices:** Consider a tree $T$ of $n$ vertices with a high depth, say, $\Theta(n)$. Intuitively, such high depths are caused by long chains of vertices, where vertices in the chain

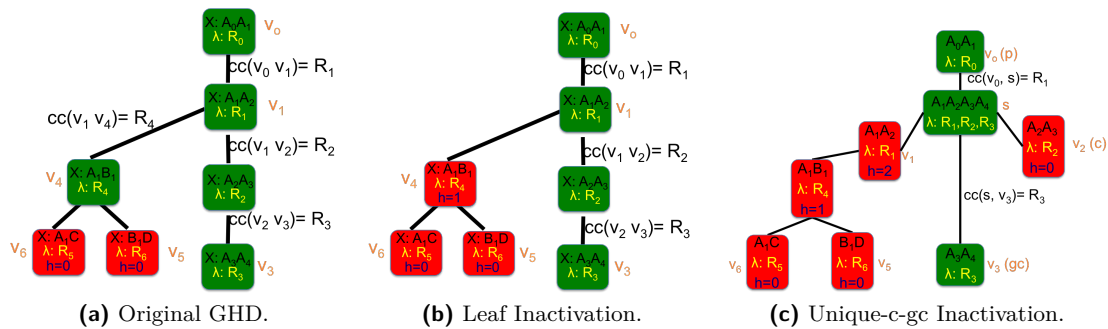

**(a)** Original GHD.   **(b)** Leaf Inactivation.   **(c)** Unique-c-gc Inactivation.

**Figure 4** Effects of leaf inactivation and unique-c-gc inactivation.

have only a single child. Log-GTA reduces the depth of high-depth GHDs by identifying and "branching out" such chains. At a high-level, Log-GTA finds a vertex $v$ with a unique child $c$ (for child), which also has unique child $gc$ (for grandchild), and puts $v$, $c$, and $gc$ under a new vertex $s$. We call vertices like $v$ *unique-c-gc* vertices.

In each iteration, Log-GTA identifies a set of nonadjacent unique-c-gc vertices and leaves of active($T'$), and inactivates them (while shortening the chains of unique-c-gc vertices). We next state an important lemma that will help bound the number of iterations of Log-GTA (proved in Appendix D.3):

▶ **Lemma 16.** *In a tree with $N$ vertices, we can find two sets $L'$ and $U'$ such that $|L'|+|U'| > \lceil \frac{N}{4} \rceil$ and vertices in $L'$ are leaves, and vertices in $U'$ are (1) unique-c-gc vertices; and (2) pairwise non-adjacent.*

## 6.2 Two Transformation Operations

We next describe the two operations that Log-GTA performs on the nodes of active($T'$).

**Leaf Inactivation:** Takes a leaf $l$ of active($T'$) and (1) sets its label to inactive; and (2) sets height($l$) to $\max\{0, \max_c\{\text{height}(c)\} + 1\}$, where $c$ is over the (inactive) children of $l$. $\chi(l)$ and $\lambda(l)$ remain the same. The common-cover between $l$ and $l$'s parent is removed.

**Unique-c-gc (And Child) Inactivation**: Let $u$ be a unique-c-gc vertex in active($T'$). Note that $u$ is not necessarily a unique-c-gc vertex in $T'$. Let $u$'s parent be $p$ (if one exists), $u$'s child be $c$, and $u$'s grandchild be $gc$. Unique-c-gc inactivation does the following:

(1) Creates a new active vertex $s$, where $\lambda(s) = cc(p,u) \cup cc(u,c) \cup cc(c,gc)$ and $\chi(s) = (\chi(p) \cap \chi(u)) \cup (\chi(u) \cap \chi(c)) \cup (\chi(c) \cap \chi(gc))$.
(2) Inactivates $u$ and $c$. Similar to leaf inactivation, sets their heights to 0 if they have no inactive children, and one plus the maximum height of their inactive children otherwise.
(3) Removes the edges $(p, u)$ and $(u, c)$ and adds an edge from $s$ to both $u$ and $c$.
(4) Adds an edge from $p$ to $s$ with $cc(p,s) = cc(p,u)$ and $s$ to $gc$ with $cc(s,gc) = cc(c,gc)$.

Figure 4b shows the effect of leaf inactivation on vertex $v_4$ of the extended GHD in Figure 4a. In the figure, green and red indicate that the vertex is active and inactive, respectively. The attributes of each $R_i$ are the $\chi$ values on the nodes that $R_i$ is assigned to. Figure 4c shows the effect of Unique-c-gc Inactivation on a unique-c-gc vertex $v_1$ from Figure 4b. We next state a key lemma about these two operations:

```
1  Input: GHD D(T, χ, λ) for hypergraph H
2  Extend D into D'(T', χ', λ') as described in Section 6.1.
3  while(there are active nodes in T')
4     Select at least ¼ of the active vertices that are either leaves L'
5                            or non-adjacent unique-c-gc vertices U'
6     Inactivate each l ∈ L', each u ∈ U' and the child of u
7  return D'
```

**Figure 5** Log-GTA.

▶ **Lemma 17.** *Assume that an extended GHD $D'(T', \chi', \lambda')$ of a hypergraph $H$ with active/inactive labels on $V(T')$, and common covers on $E(T')$ initially satisfies the following five properties:*

1. *active($T'$) is a tree.*
2. *The subtree rooted at each inactive vertex $v$ contains only inactive vertices.*
3. *The height of each inactive vertex $v$ is $v$'s correct height in $T'$.*
4. *$|cc(u, v)| \leq$ iw between any two active vertices $u$ and $v$ and does indeed cover the shared attributes of $u$ and $v$.*
5. *$D'$ is a GHD of $H$ with width at most $\max(\mathsf{w}, 3\mathsf{iw})$.*

*Performing any sequence of leaf and unique-c-gc inactivations maintains these five properties.*

We prove this lemma in Appendix D.4. We next state an immediate corollary to Lemma 17.

▶ **Corollary 18.** *Let $D(T, \chi, \lambda)$ be a GHD of a hypergraph $H$ with width $\mathsf{w}$, intersection width iw. Consider extending $D$ to GHD $D'(T', \chi', \lambda')$ with active/inactive labels, common-covers, and heights as described in Section 6.1, and then applying any sequence of leaf and unique-c-gc inactivations on $D'$. Then the resulting $D'$ is a GHD with width at most $\max(\mathsf{w}, 3\mathsf{iw})$ and the height of each inactive vertex $v$ is $v$'s actual height in $T'$.*

### 6.3  Log-GTA

Finally, we present our Log-GTA algorithm. Log-GTA takes a GHD $D$ and extends it into $D'$ by following the procedure in Section 6.1. Then, Log-GTA iteratively inactivates a set of active leaves $L'$ and nonadjacent unique-c-gc vertices $U'$ (along with the children of $U'$), which constitute at least $\frac{1}{4}$ fraction of the remaining active vertices in $T'$ by Lemma 16, until all vertices are inactive. Figure 5 shows the pseudocode of Log-GTA. Appendix F shows an example execution of Log-GTA on a width-2 and intersection width-1 GHD of $TC_{15}$ from Table 1. We next state two lemmas about Log-GTA and then prove our second main result.

▶ **Lemma 19.** *Log-GTA takes $O(\log(|V(T)|))$ iterations.*

**Proof.** Observe that both leaf inactivation and unique-c-gc inactivation decrease the number of active vertices in $T'$ by 1. In each iteration the number of active vertices decreases by a factor of $\frac{1}{4}$. Therefore the algorithm terminates in $O(\log(|V(T)|)$ iterations.  ◀

▶ **Lemma 20.** *The height of each inactive vertex $v$ is at most the iteration number at which $v$ was inactivated.*

**Proof.** By Corollary 18, the heights assigned to vertices are their correct heights in the final GHD returned. Moreover the height numbers start at 0 in the first iteration and increase by at most one in each iteration, because in each iteration, the height of each inactivated vertex $v$ is set to the maximum of $v$'s inactive children plus one. Therefore the height numbers assigned in iteration $i$ are less than $i$, completing the proof. ◀

▶ **Theorem 21** (Second Main Result). *Given any GHD $D(T, \chi, \lambda)$ with width w, intersection width iw, we can construct a GHD $D'(T', \chi', \lambda')$ where width $w' \leq \max(w, 3iw)$, $\mathrm{depth}(T') = \min\{\mathrm{depth}(T), O(\log(|V(T)|))\}$.*

**Proof.** By Corollary 18 the width of $D'$ is at most $\max(w, 3iw)$. By Lemmas 17, 19 and 20, the height of each vertex $v$ is $v$'s true height in the tree and is at most the maximum iteration number, which is $O(\log(|V(T)|))$. Therefore, the depth of $T'$ is $O(\log(|V(T)|))$. Also, the leaf and unique-c-gc inactivation operations never increase the depth of the tree, justifying that the depth of the final tree is $\min\{\mathrm{depth}(T), O(\log(|V(T)|))\}$. ◀

Theorems 15 and 21 and Lemma 7 imply the following two results:

▶ **Corollary 22.** *Given a hypergraph $H$ with $n$ hyperedges, width w, intersection width iw, we can construct a $\log(n)$ depth GHD of $H$ with width at most $\max(w, 3iw)$.*

▶ **Theorem 23.** *Any query $Q$ with width w can be executed in $O(\log(n))$ rounds and $O(n\mathrm{B}(\mathrm{IN}^{\max(w,3iw)} + \mathrm{OUT}, M))$ communication.*

We note that Corollary 22 shows that, similar to Bodlaender's and Akatov's results about log-depth TDs and HDs, a similar and stronger property also holds for GHDs. In the appendix we further show: (1) Log-GTA without any modifications recovers Bodalender's result about TDs (Appendix G.1); and (2) a modification of Log-GTA recovers Bodlaender's and Akatov's results and a weaker version of our result (Appendix G.2).

Surprisingly, if we simulate GYM on PRAM using a log-depth GHD generated by Log-GTA, we show that any bounded-width query can be evaluated in logarithmic PRAM steps using polynomial number of processors, i.e., bounded-width queries are in the complexity class NC—a result that was proven by the ACQ algorithm [15]. This is interesting in itself, since we recover this positive parallel complexity result by using only a simple variant of Yannakakis's algorithm, which has been thought to be an inherently sequential algorithm.

## 7 C-GTA (Constant-depth GHD Transformation Algorithm)

Our C-GTA algorithm is based on the following observation. For any two adjacent nodes $t_1, t_2 \in V(T)$, we can "merge" them and replace them with a new node $t \in V(T)$ and set $\chi(t) = \chi(t_1) \cup \chi(t_2)$, $\lambda(t) = \lambda(t_1) \cup \lambda(t_2)$ and set $t$'s neighbors to the union of neighbors of $t_1$ and $t_2$. As long as $t_1$ and $t_2$ were either neighbors, or both leaves with the same parent, $T$ remains a valid GHD tree after this operation. C-GTA operates as follows:
1. For each node $u$ that has an even number of leaves as children, divide $u$'s leaves into pairs and merge each pair.
2. For each node $u$ that has an odd number of leaves as children, divide the leaves into pairs and merge them, and merge the remaining leaf with $u$.
3. For each vertex $u$ that has a unique child $c$, if $c$ has an even number of leaf children, then merge $u$ and $c$.

If $T$ has $L$ leaves and a set $U$ of pairwise non-adjacent unique-c-gc nodes, then the above procedure removes at least $\frac{\max(L,U)}{2}$ nodes from $T$. We next state a combinatorial lemma to bound this quantity, which is proved in Appendix D.3.

▶ **Lemma 24.** *Suppose a tree has N nodes, L of which are leaves, and U of which are unique-c-gc nodes. Then $4L + U \geq N + 2$.*

By Lemma 24, $\max(L, U)$ is at least $n/8$. Therefore, the resulting tree $T'$ has at most $15n/16$ nodes, and width $\leq 2\mathsf{w}$. We can use this operation repeatedly to reduce the number of vertices while increasing width. We can then apply Log-GTA to get the following theorem:

▶ **Theorem 25.** *For any query Q with a width-$\mathsf{w}$, intersection width-$\mathsf{iw}$ GHD $D = (T, \chi, \lambda)$, for any i, there exists a GHD $D' = (T', \chi', \lambda')$ with width $\leq 2^i . \max(\mathsf{w}, 3\mathsf{iw})$ and depth $\leq \log((\frac{15}{16})^i n)$.*

Thus we can further trade off communication by constructing trees of even lower depth than a single invocation of Log-GTA.

## 8 Conclusions and Future Work

We have shown that by using GYM as a primitive and proving different properties of depths and widths of GHDs of queries, we can trade off communication against number of rounds of computations. We believe our approach of discovering such tradeoffs using different combinatorial properties of GHDs is a promising direction for future work. An important open area is to explore other GHD construction algorithms that output GHDs with different depths and widths. Specifically, we believe it is plausible that an algorithm can generate polynomially lower depth GHDs with twice their widths (instead of the constant depth reduction of C-GTA). We also plan to investigate the lower bounds on the communication costs of algorithms that run $O(\log(n))$ or $O(n)$ rounds.

## 9 Acknowledgements

We would like to thank the anonymous ICDT reviewers whose comments and suggestions about many parts of this paper were critical when preparing the final version of this paper.

## A   Comparison of Other Distributed BSP Models

We compare our tuple-based MR model with other distributed BSP models in literature.

|  | **Shares**$(S_n)$ | **ACQ-MR**$(S_n)$ | **GYM**$(D_{S_n})$ |
|---|---|---|---|
| **# Rounds** | 1 | $O(\log(n))$ | $O(\log(n))$ |
| **Communication** | $O(\frac{\text{IN}^{\frac{n}{2}}}{M^{\frac{n}{2}}} + \text{OUT})$ | $O(n\frac{(\text{IN}^3+\text{OUT})^2}{M})$ | $O(n\frac{(\text{IN}+\text{OUT})^2}{M})$ |

■ **Table 2** Worst-case performance of algorithms on $S_n$. $D_{S_n}$ is a $O(1)$-depth GHD of $S_n$.

**Massively Parallel Communication (MPC) [6]:**  In MPC (and its earlier version MP [24]), an algorithm picks the number of machines $p$ that it will use explicitly and in each round, each machine is allowed to receive $M = O(\frac{\text{IN}}{p^{1-\epsilon'}})$ amount of data, where $\epsilon'$ is referred to as the space exponent. Therefore the communication of each round is $C = O(pM) = O(p^{\epsilon'}\text{IN})$. There are two versions of MPC, bit-based and tuple-based, defining the cost unit of the model and what type of data can be sent between machines. Our model is similar to tuple-based MPC, where instead of fixing $p$, we fix $M$ and calculate the algorithm's communication cost $C$ in each round, which implies that the number of machines $p$ used by the algorithm in that round is $\frac{C}{M}$. However we are more flexible in that we do not necessarily require $M$ and $p$ to be related by the formula $M = O(\frac{\text{IN}}{p^{1-\epsilon'}})$ for some $\epsilon'$. In fact, for GYM, the precise relation of $p$ and $M$ will depend on IN and OUT and cannot be expressed as $M = O(\frac{\text{IN}}{p^{1-\epsilon'}})$. For example, GYM evaluates a width-w query in $O(n)$ rounds with $\frac{(\text{IN}^w+\text{OUT})^2}{M}$ communication per round. Therefore $p = \frac{(\text{IN}^w+\text{OUT})^2}{M^2}$ or $M = \frac{(\text{IN}^w+\text{OUT})}{p^{1/2}}$, which cannot be expressed as $O(\frac{\text{IN}}{p^{1-\epsilon'}})$ for some fixed $\epsilon'$; it will depend on OUT. However there are also algorithms in our model, whose memory requirement $M$ will relate to IN with a fixed space exponent $\epsilon'$ when simulated in MPC (if we do not count the writing of the final output as communication cost). For example, consider our join algorithm from Lemma 8 joining two equal sized relations $R$ and $S$. Then for any $M$ one can show that in the MPC simulation of this algorithm the space exponent is always $\epsilon' = 1/2$. For any $M$, the algorithm divides the tuples of $R$ and $S$ into $g_r = \frac{2|R|}{M}$ and $g_s = \frac{2|S|}{M}$ groups, respectively, and uses is $p = g_r \times g_s$ machines. Assuming $|R| = |S|$, $p = g_r^2$, and $\text{IN} = 2|R|$. So $M = \frac{2|R|}{g_r} = \frac{\text{IN}}{p^{1/2}}$. In fact, for all one round algorithms we know, e.g., the matrix multiplication algorithm from [2], we can analyze the algorithms equivalently in MPC and our model, i.e., we can calculate $M$, $C$ and derive $p$ in our MR model, and it will be such that $M = \frac{\text{IN}}{p^{1-\epsilon'}}$, for some fixed $\epsilon'$.

**MRC [22]:**  MRC is the earliest MapReduce model introduced. Our model is similar to MRC with two differences: (1) Instead of bits we use tuples as units of communication; (2) MRC bounds the inputs to reducers to be $O(\text{IN}^{1-\epsilon})$ and the total communication per round to be $O(\text{IN}^{2-2\epsilon})$ for some constant $\epsilon \in (0,1)$, whereas we allow these costs to take any value.

**Model of MapReduce in Reference [12]:**  Our model is very similar to the MR model in reference [12], except we use tuples instead of bits as our cost unit and the model there defines a complex runtime metric, similar to parallel models like LogP [9], depending on latency and bandwidth of the communication network. We use the simpler cost metrics of rounds, memory, and total communication as these metrics reflect the fundamental tradeoffs in the evaluation of conjunctive queries that will hold independently of the underlying network.

|  | **Shares**($TC_n$) | **ACQ-MR**($TC_n$) | **GYM(Log-GTA**($D_{TC_n}$)) | **GYM**($D_{TC_n}$) |
|---|---|---|---|---|
| **# Rounds** | 1 | $O(\log(n))$ | $O(\log(n))$ | $O(n)$ |
| **Communication** | $O(\frac{\text{IN}^{\frac{n}{6}}}{M^{\frac{n}{6}}} + \text{OUT})$ | $O(n\frac{(\text{IN}^6+\text{OUT})^2}{M})$ | $O(n\frac{(\text{IN}^3+\text{OUT})^2}{M})$ | $O(n\frac{(\text{IN}^2+\text{OUT})^2}{M})$ |

■ **Table 3** Worst-case performance of algorithms on $TC_n$. $D_{TC_n}$ is a $\theta(n)$-depth GHD of $TC_n$.

## B  GYM vs ACQ-MR

The ACQ algorithm [15] is the most efficient known $O(\log(n))$-step PRAM algorithm for computing bounded-width queries. We refer the reader to reference [15] for the details of the algorithm. The algorithm takes as input an acyclic query and first runs an algorithm called *FULL-REDUCER* to remove tuples from the input relations that do not contribute to the final output. Irrespective of the depth of the input query, FULL-REDUCER runs for $\Theta(\log(n))$ steps of PRAM. In each step, similar to our Log-GTA algorithm, FULL-REDUCER performs one of two operations called *shunt(mark)* or *r-shunt*, which effectively contract a join tree. The shunt operations perform multiple semijoins and joins in parallel, where each join operation is between exactly three base relations, and some of the semijoins are between two intermediate relations and some are between base relations. After FULL-REDUCER finishes, the input relations only contain tuples that contribute to the final output. The algorithm performs $\Theta(\log(n))$ PRAM steps of another operation called *shunt*, which are between semijoined base relations or intermediate relations that join three base relations. The FULL-REDUCER algorithm is analogous to combining our Log-GTA algorithm with the upward and downward semijoin sub-phases of Yannakakis's algorithm. And the second step of ACQ is analogous to the join phase of Yannakakis's algorithm. The difference is that FULL-REDUCER on any acyclic query runs for $\Theta(\log(n))$ PRAM steps, and will always join three of the base input relations as intermediate relations. We can use our basic operations from Lemmas 8 and 10 to perform the join and semijoin operations of ACQ in MR. Moreover for cyclic queries, we can use the materialization step of GYM to generalize ACQ. We call this algorithm *ACQ-MR*.

ACQ-MR evaluates a width-w query $Q$ in $\Theta(\log(n))$ rounds and $O(nB(\text{IN}^{3\text{w}} + \text{OUT}, M))$ communication because during its shunt operations ACQ always joins exactly three base relations to create its intermediate relations, which will be of size $O(\text{IN}^{3w})$ in the worst-case. Let *GYM(Log-GTA(D))* refer to our combined algorithm that first runs our Log-GTA algorithm on a GHD D of $Q$ and constructs a new GHD $D'$, and then runs GYM with $D'$ to evaluate $Q$. Since GYM(Log-GTA) also runs in $\Theta(\log(n))$ rounds (note that $D'$ will have a depth of at most $O(\log(n))$), and has a communication cost of at most $O(nB(\text{IN}^{3\text{w}} + \text{OUT}, M))$, GYM(Log-GTA) always matches ACQ-MR's performance. As we next show, sometimes GYM or GYM(Log-GTA) can outperform ACQ-MR. Tables 2 and 3 summarize the two cases we consider below. In the tables, for completeness, we also compare our algorithms and ACQ-MR with the Shares algorithm, which is the best known one-round algorithm.

1. If $Q$ has low-depth width-w GHDs, such as the $S_n$ query, GYM outperforms ACQ-MR in communication cost while using a comparable number of rounds (Table 2).
2. If $Q$ has high-depth, say of $\Theta(n)$-depth, GHDs but has an intersection width (see Section 3) that is strictly lower than its width, such as the $TC_n$ query, then: (1) GYM outperforms ACQ-MR in terms of communication but executes for a larger number of rounds; and (2) GYM(Log-GTA) also incurs less communication cost than ACQ-MR (but more than GYM), while using a comparable number of rounds (Table 3).

## C  Discussion on Skew

Skew in a database input refers to variation in the frequency of different attribute values. A relation without skew would be one without heavy hitter values. This has been formalized in reference [6] through matching databases. Let $N$ be the size of a universe of values that can appear in the attributes of relations. In matching databases, each attribute of each relation $R_i$ of size $|R_i|$ is a subset of a permutation of integers from 1 to $N$ for some $N \geq |R_i|$. Therefore, in a connected query, any pairwise join of relations $R$ and $S$ will generate an intermediate relation that is at most size $\min(|R|, |S|)$. This is because, each tuple in $R$ can match at most one tuple in $S$ (and vice versa). There are two other advantages of matching databases that we can use to improve the communication cost and round costs of GYM:

**Improvement on Communication:** Instead of using the join algorithm in Lemma 8 we can simply send each tuple $t$ of $R$ and $S$ to a reducer by hashing $t$ on the common attributes of $R$ and $S$ into $\max(|R|, |S|)^2$ reducers. For example, if $R = (A, B)$ and $S = (B, C)$ we can hash each tuple on attribute $B$, without any replication. Because the columns of each relation is a permutation, w.h.p. each reducer would get $O(1)$ tuples. This would be a one round algorithm with a communication cost of $|R| + |S|$, improving the $O(\mathrm{B}(|R| + |S|, M))$ communication cost of the binary join algorithm in Lemma 8. In general, for a width-w query Q, even if the input GHD $D(T, \chi, \lambda)$ of $Q$ has nodes in $V(T)$ that are Cartesian products, then GYM's cost over matching databases would still be $O(n(\mathrm{IN}^{\mathsf{w}} + \mathrm{OUT}))$, instead of $O(n\mathrm{B}(\mathrm{IN}^{\mathsf{w}} + \mathrm{OUT}, M))$.

**Improvement on Number of Rounds:** Recall that after the materialization stage GYM produces an acyclic query $Q'$ and runs DYM-d's join phase (Section 4.3) on $Q'$, which takes $O(\log(n))$ rounds. In each round, DYM-d runs multiple binary joins in parallel. Consider w.l.o.g. an acyclic query $Q'$ and its input GHD $D = (T, \chi, \lambda)$. Suppose further that a relation $R$ in $D$ has $k$ children relations $S_1, \ldots, S_k$. In order to join $R$ with each $S_i$, DYM-d runs for $\log(k)$ rounds, where in each round it evaluates multiple binary joins in parallel. Over matching databases, we can perform this join in two rounds. Since we focus on connected queries, we know that $R$ has a common attribute with each $S_i$. In the first round, we join $R$ with each $S_i$ by hashing $R$ and $S_i$ on their common attributes to $\max(|R|, |S_i|)^2$ reducers, ensuring that each reducer gets $O(1)$ tuples w.h.p. This generates $k$ intermediate tables $T_i$. Note each $T_i$ is of size at most $|R|$ because the relations are matching. Then we join all $T_i$ in another round by hashing each table on all of the attributes of $R$ (which each $T_i$ contains) to $|R|^2$ reducers, guaranteeing that each reducer gets $O(k)$ tuples. The same optimization can be done for the semijoin phase, decreasing the number of rounds from $O(\mathsf{d} + \log(n))$ to $O(\mathsf{d})$. For example, on a matching database, we could compute the join of $S_n$ in $O(1)$ rounds, as opposed to the $O(\log(n))$ rounds we execute on arbitrary databases.

## D  Proofs of Lemmas and Theorems

### D.1  Lemma 7

For any tree $T$, let $L(T)$ denote the set of leaves of $T$, $M(T)$ denote the set of vertices of $T$ with degree 2, and $N(T)$ denote the remaining vertices of $T$ (i.e. those with degree $\geq 3$). Thus $L(T), M(T), N(T)$ are pairwise disjoint and $V(T) = L(T) \cup M(T) \cup N(T)$. Moreover, in a tree $T$, $|E(T)| = |V(T) - 1|$, and $\sum_{t \in T} \mathsf{degree}(t) = 2|E(T)|$. Since $\mathsf{degree}(t) \geq 3$ for all $t \in N(T)$, we must have $|N(T)| \leq 2|E(T)|/3 \leq 2|V(T)|/3$. Thus $|L(T)| + |M(T)| \geq |V(T)|/3$. We call a GHD $D' = (T', \chi', \lambda')$ of query with hypergraph $H = (V(H), E(H))$ *minimal* if for every vertex $t \in L(T) \cup M(T)$, there exists a hyperedge $e_t \in E(H)$ such that $e_t \subseteq \chi(t)$ and $e_t \subsetneq \chi(t')$ for every $t' \neq t$ in $V(T)$. That is, for every vertex with degree $\leq 2$, there must be a hyperedge that is covered by that vertex alone.

We first show the following. Consider any GHD $D$ of $H$ and let $n$ be $|E(H)|$. Then $D$ can be turned into a minimal GHD $D'$ of $H$ with at most $3n$ tree nodes, without increasing its width, intersection width, or depth. Then we show how to turn $D'$ into a complete GHD $D''$ of $H$ without affecting $D'$'s width and intersection width and by increasing its depth and the number of tree nodes by at most 1 and $n$, respectively.

A GHD $D = (T, \chi, \lambda)$ that is not minimal can be made minimal as follows: When a leaf $t \in L(T)$ has no hyperedge that is only covered by $t$, we simply delete $t$. Similarly, when a $t \in M(T)$ has no hyperedge that is only covered by $t$, we delete $t$ and add an edge between its two neighbors. We note that these operations do not increase the width or depth of $T$. These operations also do not increase the intersection width of the query and violate any of the GHD properties of D. To see this, we only need to consider deleting a vertex $t \in M(T)$. Call one of $t$'s neighbors $p$, for parent, and the other $c$, for child. Notice that by the 2nd property of GHDs, $c$ and $p$ cannot share an attribute that they do not already share with $t$. Therefore removing $t$ and connecting $p$ and $c$ will maintain the 2nd property of GHDs. It also does not violate the 1st and 3rd properties of GHDs. Moreover, removing $t$ will not affect the intersection width of $D$ either. To see this, note that any set of hyperedges that cover the common attributes of $c$ and $t$ will also cover the common attributes of $c$ and $p$ after $t$ is removed.

We perform these operations until every remaining $t \in L(T) \cup M(T)$ has a hyperedge $e$ that is only covered by it. Thus given a width-w, intersection width-iw GHD $D = (T, \chi, \lambda)$ of depth d, we can get a minimal GHD $D' = (T', \chi', \lambda')$ that has width $\leq w$, intersection with $\leq$ iw, and depth $\leq$ d. We next show that $|V(T')| \leq 3n$. We argued earlier that $|L(T')| + |M(T')| \geq |V(T')|/3$, and there is at least one edge that is covered uniquely by each $t \in L(T') \cup M(T')$. Thus we must have $|E(H)| \geq |L(T')| + |M(T')| \geq |V(T')|/3$. Thus $|V(T')| \leq 3|E(H)| = 3n$.

Finally, we observe that if GHD $D'$ is not complete, we can modify it into a complete GHD $D''$ by adding at most $n$ new leaves. For each hyperedge $e \in E(H)$ that does not appear in the $\lambda$-label of any vertex, we assign a new leaf $l$ with $\lambda(l) = \chi(l) = e$ as a child of any vertex $v \in T$ that contains $e$ in $\chi(v)$. Note that one such $v$ has to exist by definition of a GHD. This operation does not affect the width or intersection width of $D'$, increases the depth of $D'$ by at most 1, and the size of $T'$ by at most $n$. Therefore, given a GHD $D = (T, \chi, \lambda)$ of a hypergraph $H$, with width w, intersection width iw and depth d, we can turn it into a minimal and complete GHD $D'' = (T'', \chi'', \lambda'')$ of $H$ with width $\leq$ w, intersection width $\leq$ iw, depth $\leq d + 1$, and $|V(T'')| \leq 4n$.

## D.2 Lemma 13

Steps (1) and (2) of each recursive call can be performed in $O(1)$ rounds, in parallel for all leaves. So we only need to prove that the number of recursive calls is $O(\mathsf{d} + \log(n))$. For any tree $T$, let $X(T) = \sum_{l \in L(T)} 2^{\mathsf{d}(l)}$, where $L(T)$ denotes the leaves in $T$ and $\mathsf{d}(l)$ is the depth of leaf $l$. Then, $X$ of the original join tree $T$ is at most $n 2^{\mathsf{d}}$, as there are at most $n$ leaves, with depth at most d each. Now consider what happens to $X$ of the tree in each recursive call. Each leaf $l$ is in either $L_1$ or $L_2$. If it is in $L_1$, it gets deleted. If $l$'s parent has no other children, then the parent becomes a new leaf, of depth $\mathsf{d}(l) - 1$. Thus the $2^{\mathsf{d}(l)}$ term in $X$ is at least halved for all leaves in $L_1$. On the other hand, if $l_1, l_2$ form a pair in $L_2$, then one of them gets deleted, while the other stays at the same depth. Thus the $2^{\mathsf{d}(l_1)} + 2^{\mathsf{d}(l_2)}$ term also gets halved. For a triple in $L_2$, the term becomes one-third. Thus $X$ reduces by at least half in each recursive call. The recursion terminates when $T$ is a single node, i.e. when $X(T) = 1$. Since its starting value is at most $n2^d$, the number of recursive calls is $O(\log(n2^{\mathsf{d}})) = O(\mathsf{d} + \log(n))$.

## D.3   Lemmas 16 and 24

We first prove Lemma 24 and then prove Lemma 16. First we state and prove another lemma that we will also need in the proof of Lemma 16.

▶ **Lemma 26.** *If a tree has $U$ unique-c-gc nodes, we can select at least $\lceil \frac{U}{2} \rceil$ of them that are not pairwise adjacent to each other.*

**Proof.** We process the tree top-down and whenever we see a unique-c-gc node we select it and mark its unique child as forbidden (i.e., we cannot select it afterwards). Note that the marked node may or may not be a unique-c-gc node itself. As each selected unique-c-gc node marks at most one other node as forbidden, this process takes at least half of the unique-c-gc nodes, and because we mark the immediate child of each selected unique-c-gc node as forbidden, the selected nodes are pairwise non-adjacent.   ◀

We next prove Lemma 24. The proof is an induction on the height $h$ of the tree.

BASIS: If h=0, then the root is the only node in the tree and is a leaf. Therefore, $4L + U = 4 \geq 1 + 2 = 3$. If $h = 1$, then the tree is a root plus $N - 1$ children of the root, all of which are leaves. Thus, $L = N - 1$, and $U = 0$. We must verify $4(N - 1) + 0 \geq N + 2$, or $3N \geq 6$. Since $N$ is necessarily larger than 2 for any tree of height at least 1, we may conclude the bases.

INDUCTION: Now, assume $h \geq 1$. There are three cases to consider: *Case 1*: The root has a single child $c$ and $c$ has a single child $gc$. Then the root is a unique-c-gc node and the tree rooted at $c$ has L leaves, $U - 1$ unique-c-gc nodes, and a total of $N - 1$ nodes. By the induction hypothesis, $4L + U - 1 \geq (N - 1) + 2$, or $4L + U \geq N + 2$, which completes the induction in this case. *Case 2*: The root has a single child, which has $k \geq 2$ children $c_1, \ldots, c_k$. Let the subtree rooted at $c_i$ have $L_i$ leaves, $U_i$ unique-c-gc nodes, and $N_i$ nodes. By the inductive hypothesis, $4L_i + U_i \geq N_i + 2$. Summing over all $i$ we get $4L + U \geq (N - 2) + 2k$. Since $k \geq 2$, we conclude $4L + U \geq N + 2$, which completes the induction in this case. *Case 3*: The root has $k \geq 2$ children $c_1, \ldots, c_k$. Similarly, if the subtree rooted at $c_i$ has $L_i$ leaves, $U_i$ unique-c-gc nodes, and $N_i$ nodes and we sum over all $i$, we get $4L + U \geq (N - 1) + 2k$. Since $k \geq 2$, we again conclude that $4L + U \geq N + 2$, which completes the proof.

Using Lemmas 26 and 24, we can now prove Lemma 16. Let the tree have $N$ nodes, $L$ leaves, and $U$ unique-c-gc nodes. Lemma 24 says that $4L + U \geq N + 2$. By Lemma 26, we can select at least $\lceil U/2 \rceil$ of the unique-c-gc nodes. Since we may also select all leaves, and $L + \lceil U/2 \rceil \geq L + U/4 \geq N/4 + 2/4 \geq N/4$. Since the left side is an integer, we can conclude that we can select at least $\lceil N/4 \rceil$ of the vertices that can be partitioned into leaves and a set of pairwise non-adjacent unique-c-gc vertices, completing the proof.

## D.4   Lemma 17

Let $D'(T', \chi', \lambda')$ be a GHD that satisfies these five properties. First, consider inactivating an active leaf $l$ of $D'$.
1. For property (1), we observe that inactivating $l$ removes a leaf of active($T'$), so active($T'$) remains a tree after the operation.

2. For property (2), we only need to consider the subtree $S_l$ rooted at $l$. Observe that none of $l$'s children can be active, since this would contradict that $l$ is a leaf of active($T'$). In addition, none of $l$'s other descendants can be active because then the subtree rooted at one of $l$'s inactive children would contain an active vertex. This would contradict the assumption that initially all subtrees rooted at inactive vertices contained only inactive vertices.

3. For property (3), notice that the height that is assigned to $l$ is 0 if it has no children, which is its correct height in $T'$. Otherwise, $l$'s height is one plus the maximum of the heights of $l$'s children, which is also its correct height in $T'$ since all of $l$'s children are inactive and have correct heights by assumption.

4. Properties (4) and (5) hold trivially as leaf inactivation does not affect the common-covers, $\chi$, and $\lambda$ values and by assumption their properties hold in D$'$.

Now let's consider a unique-c-gc inactivation operation.

1. Property (1) holds because by definition $u$ has one active child $c$ and $c$ also has one active child $gc$. So $u$ and $c$ are part of a chain of active($T'$). The unique-c-gc inactivation merges $u$ and $c$ together into another active vertex $s$ on this chain without affecting the acyclicity or connectedness of active($T'$). Notice that we also add two edges from $s$ to $u$ and $c$. But $u$ and $c$ are now inactive.

2. For property (2) observe that the only two subtrees we need to consider are the subtrees rooted at $u$ and $c$, which we call $S_u$ and $S_c$, respectively. Notice that all of the edges that go down the tree from $u$ and $c$ after removing $(u, c)$ and $(c, gc)$ were to inactive vertices. Therefore, by the same argument we did for leaf elimination, both $S_u$ and $S_c$ have to consist of only inactive vertices.

3. We assign heights to $u$ and $c$ in the same way as we assigned the height of an inactivated leaf. The exact same argument we made for leaf elimination proves that $u$ and $c$ get assigned their correct heights in $T'$.

4. We need to consider two common covers: $cc(p, s)$, which is assigned $cc(p, u)$, and $cc(s, gc)$, which is assigned $cc(c, gc)$. The sizes of $cc(p, s)$ and $cc(s, gc)$ are at most iw because the sizes of $cc(p, u)$ and $cc(c, gc)$ are at most iw initially by assumption. We next prove that $cc(p, s)$ indeed covers the common attributes between $p$ and $s$. The proof for $cc(s, gc)$ is symmetric and omitted. Notice that since $\chi(s) = (\chi(p) \cap \chi(u)) \cup (\chi(u) \cap \chi(c)) \cup (\chi(c) \cap \chi(gc))$, $\chi(s) \cap \chi(p)$ is exactly equal to $\chi(p) \cap \chi(u)$. This follows because $\chi(s) \cap \chi(p)$ is $(\chi(p) \cap \chi(u)) \cup (\chi(p) \cap \chi(u) \cap \chi(c)) \cup (\chi(p) \cap \chi(c) \cap \chi(gc))$, which are equal to $(\chi(p) \cap \chi(u)) \cup (\chi(p) \cap \chi(c) \cap \chi(gc))$. This is the set $(\chi(p) \cap \chi(u))$ union the set of attributes that $p$, $c$ and $gc$ share together. However this is exactly equal to $(\chi(p) \cap \chi(u))$ because $p$ cannot share an attribute with $c$ or $gc$, say $A_i$, that it does not already share with $u$, as this would contradict that the subtree containing $A_i$ in $D'$ is connected (and therefore contradicting that $D'$ is a GHD). By assumption $cc(p, u)$ covers $\chi(p) \cap \chi(u)$. Therefore $cc(p, s)$, which includes $cc(p, u)$, covers $\chi(p) \cap \chi(s)$.

5. For property (5), we need to prove that the three properties of GHDs hold and also verify that the width of the modified $D'$ is at most max(w, 3iw).
    - **1st property of GHDs:** Initially, before the unique-c-gc inactivation, each $e \in E(H)$ is covered by $\lambda(v)$ of some $v \in V(T')$. Since unique-c-gc does not delete any vertex in $V(T')$ (it just inactivates $u$ and $c$), this property still holds.
    - **2nd property of GHDs:** We need to verify that for each attribute $X$, the vertices that contain $X$ must be connected. It is enough to verify that all attributes among $p$, $s$, $u$, $c$, and $gc$ are locally connected, since other parts of $T'$ remain unchanged. We need to consider all possible breaks in connectedness between $p$, $u$, $c$, and $gc$ introduced by

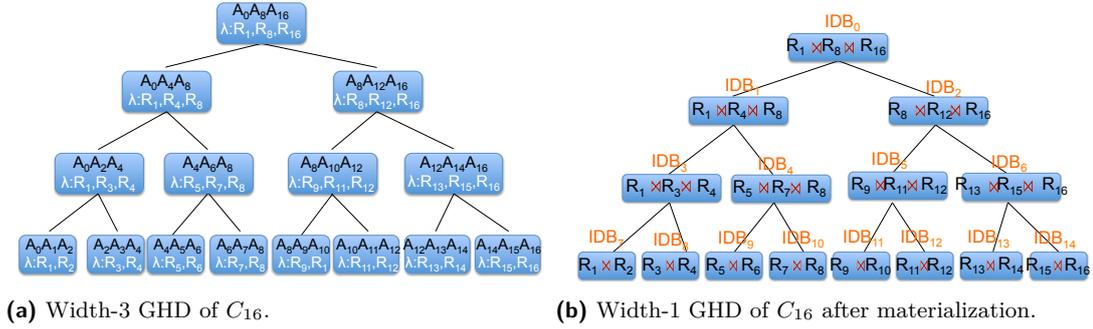

**(a)** Width-3 GHD of $C_{16}$.

**(b)** Width-1 GHD of $C_{16}$ after materialization.

**Figure 6** GHDs of $C_{16}$ at various steps of GYM(Log-GTA).

the insertion of $s$. The proof of each combination is the same. We only show the proof for attributes between $p$ and $gc$. Consider any attribute $X \in \chi(p) \cap \chi(gc)$. Then, since the initial $D'$ was a valid GHD, $X$ must have been in $\chi(u)$ (and also $\chi(c)$). Then $\chi(s)$ also includes $X$ because $\chi(s)$ includes $\chi(p) \cap \chi(u)$, proving that the vertices of $X$ are locally connected among $p$, $s$, $u$, $c$, and $gc$.

- **3rd property of GHDs:** We need to verify that for each vertex $v$, $\chi(v) \subseteq \cup \lambda(v)$. The unique-c-gc inactivation only inserts the vertex $s$, and by assumption $\chi(s)$ is the union of three intersections, each of which is covered (respectively) by the three common-covers that comprise $\lambda(s)$.
- **Width of the modified GHD:** Again by assumption, the sizes of each common cover in $\lambda(s)$ is at most iw, therefore $|\lambda(s)|$ is at most 3iw, showing that the width of the GHD is still at most max(w, 3iw).

## E  Example Execution of GYM

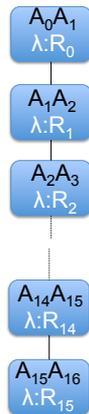

**Figure 7** Width-1 GHD of $C_{16}$.

We now describe how to compute an example query with GYM. Consider the following chain query $C_{16}: R_1(A_0, A_1) \bowtie R_2(A_1, A_2) \bowtie \cdots \bowtie R_{16}(A_{15}, A_{16})$. Figure 6a shows a width-3 GHD of this query. GYM on this GHD would first compute the IDBs in each vertex of Figure 6a. The materialized GHD, shown in Figure 6b, is now a width-1 GHD over the IDBs and therefore the join over the IDBs is acyclic. Then the algorithm simply executes DYM-d on the GHD of Figure 6b to compute the final output. Let $c$ be the (constant) number of rounds to process the semijoin of two relations. Overall the algorithm would take $12c + 6$ rounds and $O(\text{B}(\text{IN}^3 + \text{OUT}, M))$ communication cost. For comparison, Figure 7 shows a width-1 GHD of the original chain query. Executing GYM directly on this GHD would take $32c + 16$ rounds and $O(\text{B}(\text{IN} + \text{OUT}, M))$ communication cost.

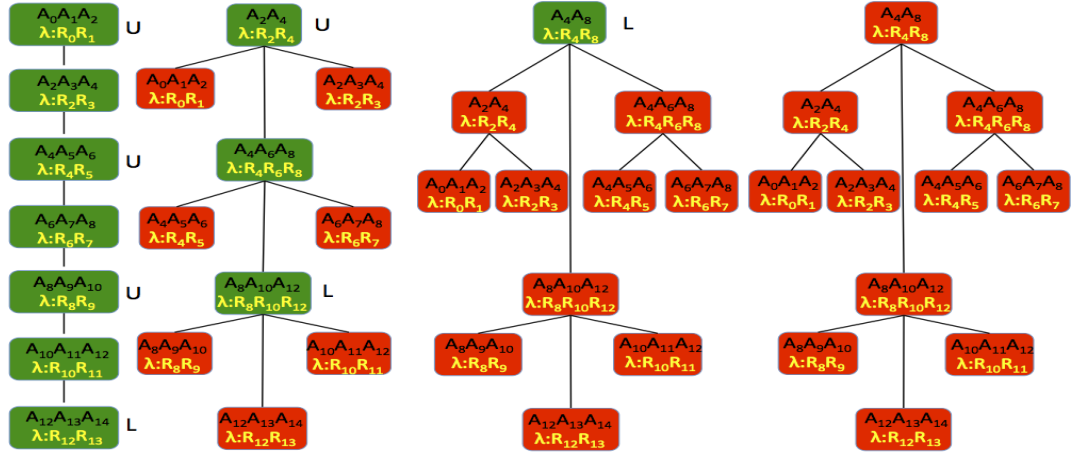

**Figure 8** Log-GTA simulation.

## F   Example Execution of Log-GTA

Figure 8 shows a simulation of Log-GTA on the width-2 and intersection width-1 GHD of $TC_{15}$ from Table 1. In the figure, Log-GTA produces a width-3 GHD with depth 2. We label the selected leaves and unique-c-gc vertices with L and U respectively, and omit the common-cover labels. In moving from the first (left most) GHD to the second GHD, Log-GTA inactivates the leaf at the bottom and the 3 unique-c-gc vertices labeled as U. Then by inactivating one leaf and one unique-c-gc vertex, the algorithm generates the third GHD. Finally by inactivating the only active vertex in the GHD, the algorithm produces the fourth (right most) GHD.

## G   Recovering and Improving Akatov's and Bodlaender's Results

### G.1   Proof that Log-GTA Recovers Bodlaender's Result

A TD $D = (T, \chi)$ of a hypergraph $H$ is a decomposition that satisfies only the first two properties of GHDs: (1) $T(V(T), E(T))$ is a tree; and (2) $\chi : V(T) \to 2^{V(H)}$ is a function associating a set of vertices $\chi(t) \subseteq V(H)$ to each vertex $t$ of $T$. Therefore, by definition each GHD is a TD. The *treewidth* of a TD is $\max_{t \in V(T)}\{|\chi(t)|\} - 1$. The treewidth of a hypergraph is the minimum treewidth of any of its TDs.

Since each GHD is also a TD, Log-GTA also transforms a TD into another TD. We will prove that given a TD with treewidth tw (that may not necessarily be a GHD), the final output of Log-GTA is a log-depth TD with $3\mathsf{tw}+2$. Bodlaender [7] had proved that log-depth TDs with $3\mathsf{tw}+2$ exist for hypergraphs. Therefore our result recovers Bodlaender's results.

We first extend Lemma 17:

▶ **Lemma 27.** *In addition to the five properties stated in Lemma 17, assume the initial TD $D'(T', \chi')$ satisfies the following property:*

6. $|\chi(u) \cap \chi(v)| \leq k$ *between any two active vertices $u$ and $v$.*

*Then performing any sequence of leaf and unique-c-gc inactivations maintains this property.*

**Proof.** Let $D'(T', \chi')$ be a TD that satisfies this property. Inactivating an active leaf $l$ of $D'$ does not affect the $\chi$ values and by assumption property (6) holds in $D'$. Now, let's consider a unique-c-gc inactivation operation. We only need to look at $|\chi(p) \cap \chi(s)|$ and $|\chi(s) \cap \chi(gc)|$. We argued in the proof of property (4) that $\chi(p) \cap \chi(s) = \chi(p) \cap \chi(u)$, which by assumption has a size of at most $k$. Similarly, $\chi(s) \cap \chi(gc)$ is exactly equal to $\chi(c) \cap \chi(gc)$, since $gc$ cannot share an attribute with $u$ or $p$ that it does not already with $c$. Again, by assumption $|\chi(c) \cap \chi(gc)|$ is at most k. ◀

Next we state a modified version of Corollary 18:

▶ **Corollary 28.** *Let $D(T, \chi)$ be a TD with treewidth* tw. *Consider extending $D$ to TD $D'(T', \chi')$ with active/inactive labels, common-covers, and heights as described in Section 6.1, and then applying any sequence of leaf and unique-c-gc inactivations on $D'$. Then the resulting $D'$ is a TD with treewidth at most $3\mathsf{tw} + 2$ and the height of each inactive vertex $v$ is $v$'s actual height in $T'$.*

**Proof.** Assume initially that $|\chi(u) \cap \chi(v)| \leq k$ between any two active vertices $u$ and $v$ in $D'$. Observe that a series of leaf and unique-c-gc inactivations only introduces new $s$ vertices through unique-c-gc inactivations. The size of $\chi(s)$ for any $s$ is at most $3k$ because $\chi(s)$ is the union of three sets, each with size at most $k$ by assumption. Recall that by definition $\mathsf{tw} = \max_{t \in V(T)}\{|\chi(t)|\} - 1$, which is at least $k - 1$. Therefore $3k$ is at most $3\mathsf{tw} + 3$, which implies that the treewidth of $D'$ is at most $3\mathsf{tw} + 2$. ◀

Now an immediate corollary to Corollary 28 and the fact that Log-GTA outputs log depth TDs (Theorem 21) recovers Bodlaender's result, which we state next:

▶ **Corollary 29.** *Given a hypergraph $H$ with $n$ hyperedges, treewidth* tw, *we can construct a $\log(n)$ depth TD of $H$ with treewidth at most $3\mathsf{tw} + 2$.*

## G.2   Log-GTA′: Recovering Both Akatov's and Bodlaender's Results

In order to recover both Akatov's and Bodlaender's results, we make two changes to the unique-c-gc inactivation operation. The leaf inactivation operation and the remainder of the algorithm remains exactly the same. We call this algorithm Log-GTA′.

1. In Log-GTA, the common-covers are essentially $\lambda$ values that are assigned to each active edge of the input $D'$. Instead, just as is done to the vertices of GHDs, we assign both $\lambda$ values and $\chi$ values to each active edge $(u, v)$ of $D'$. To help differentiate these values from $\lambda$ and $\chi$ values assigned to vertices, we will denote them in capital letters, as $\Lambda(u, v)$ and $X(u, v)$. Initially, for each edge $(u, v)$, $\Lambda(u, v) = \lambda(v)$, and $X(u, v) = \chi(v)$, i.e., copies of the child node $v$'s labels.

2. In each unique-c-gc inactivation, the new vertex $s$ is assigned $\lambda(s) = \Lambda(p, u) \cup \Lambda(u, c) \cup \Lambda(c, gc)$, $\chi(s) = X(p, u) \cup X(u, c) \cup X(c, gc)$. The new edge $(p, s)$ is assigned $\Lambda(p, s) = \Lambda(p, u)$, and $X(p, s) = X(p, u)$. Finally, the new edge $(s, gc)$ is assigned $\Lambda(s, gc) = \Lambda(c, gc)$, and $X(s, gc) = X(c, gc)$.

We next state a theorem whose proof is given momentarily in Appendix G.3.

▶ **Theorem 30.** *Given any GHD $D(T, \chi, \lambda)$ with width* w, *and treewidth* tw, *Log-GTA′ constructs a GHD $D'(T', \chi', \lambda')$ where width $\mathsf{w}' \leq 3\mathsf{w}$, treewidth $\mathsf{tw}' \leq 3\mathsf{tw} + 2$, $\mathrm{depth}(T') = \min\{\mathrm{depth}(T), O(\log(|V(T)|))\}$, and $|V(T')| \leq 2|V(T)|$. In addition, if the initial GHD is an HD, then $D'$ is also an HD.*

Consider a hypergraph $H$ with $n$ hyperedges and (generalized hypertree) width w, hypertree width hw, and treewidth tw. Three corollaries to Theorem 30, and Lemmas 19 and 20, that state that Log-GTA (and Log-GTA$'$) take a logarithmic number of iterations are:

1. A GHD of $H$ with $O(\log(n))$-depth and width $\leq 3\mathsf{w}$ exist.
2. An HD of $H$ with $O(\log(n))$-depth and width $\leq 3\mathsf{hw}$ exist.
3. A TD of $H$ with $O(\log(n))$-depth and treewidth $\leq 3\mathsf{tw} + 2$ exist.

Therefore, using a single construction, Log-GTA$'$ recovers a weaker version of our result about GHDs (instead of $w' \leq \max(\mathsf{w}, 3\mathsf{iw})$, we get $w' \leq 3\mathsf{w}$), and recovers both Bodlaender's and Akatov's results.

## G.3  Proof of Theorem 30

We start by proving that a slight modification of Lemma 17 holds about a series of leaf inactivation and the modified unique-c-gc inactivation (modified-unique-c-gc inactivation hereon) operations applied to a GHD $D$ that has been extended with the new $\Lambda$ and $X$ values on the edges described in Section G.2.

▶ **Lemma 31.** *Assume an extended GHD $D'(T', \chi', \lambda')$ of a hypergraph H with active/inactive labels on $V(T')$, and common cover labels on $E(T')$ initially satisfies the following six properties:*

1. *The active($T'$) is a tree.*
2. *The subtree rooted at each inactive vertex $v$ contains only inactive vertices.*
3. *The height of each inactive vertex $v$ is $v$'s correct height in $T'$.*
4. *For any two adjacent active vertices $u$ and $v$: (a) $|\Lambda(u, v)| \leq \mathsf{w}$; (b) $\Lambda(u,v) \supseteq X(u,v)$; (c) $X(u,v) \supseteq \chi(u) \cap \chi(v)$; and (d) $X(u,v) \subseteq \chi(v)$.*
5. *$|\chi(u) \cap \chi(v)| \leq k$ between any two active vertices $u$ and $v$.*
6. *$D'$ is a GHD with width at most $3\mathsf{w}$.*

*Then performing any sequence of leaf and the modified-unique-c-gc inactivations maintains these six properties. In addition, if the initial GHD is an HD, i.e., satisfies the descendant property (recall Section 3), then the resulting $D'$ is also an HD.*

**Proof.** We first prove that leaf and unique-c-gc operations maintain the six property for GHDs. We then prove that if $D'$ is an HD, then HD's descendant property is also maintained:

**GHD Properties:** The proof that inactivating a leaf vertex maintains these properties except property (4) is the same as the proof provided for leaf inactivation in Lemma 17. Leaf inactivation also maintains property (4) because it does not affect the $\Lambda$ and $X$ values on the edges or vertices. The proof that modified-unique-c-gc inactivation maintasin properties (1), (2), and (3) are the same as the proof provided for leaf inactivation in Lemma 17. Below we prove that modified-unique-c-gc inactivation also maintains properties (4), (5), and (6).
4. We need to consider two new active edges $(p, s)$ and $(s, gc)$. We provide the proof for $(p, s)$. The proof for $(s, gc)$ is similar and omitted.
  (a) $\Lambda(p, s) = \Lambda(p, u)$, which by assumption has size at most $\mathsf{w}$.
  (b) $\Lambda(p, s) = \Lambda(p, u) \supseteq X(p, u) = X(p, s)$.
  (c) We prove that $X(p, s) \supseteq \chi(p) \cap \chi(s)$. Notice that $\chi(s) = X(p, u) \cup X(u, c) \cup X(c, gc)$ and by assumption $X(p, u) \subseteq \chi(u)$, and $X(u, c) \subseteq \chi(c)$, and $X(c, gc) \subseteq \chi(gc)$. Since $p$ cannot share an attribute with $c$ and $gc$ that it does not already share with $u$, $\chi(p) \cap \chi(s) \subseteq \chi(p) \cap \chi(u)$ and by assumption $X(p, s) = X(p, u) \supseteq \chi(p) \cap \chi(u)$.

(d) $X(p,s) \subseteq \chi(s)$ because by construction $X(p,s) = X(p,u)$ and $\chi(s) = X(p,u) \cup X(u,c) \cup X(c,gc)$.

5. For property (5), we only need to look at $|\chi(p) \cap \chi(s)|$ and $|\chi(s) \cap \chi(gc)|$. We argued in the proof of property (4-c) that $\chi(p) \cap \chi(s) \subseteq \chi(p) \cap \chi(u)$, which by assumption has a size of at most $k$. The proof that $|\chi(s) \cap \chi(gc)| \leq k$ is similar and omitted.

6. For property (6), we need to prove that the three properties of GHDs hold and also verify that the width of the modified $D'$ is at most 3w.
   - **1st property of GHDs:** We proved in the proof of Lemma 17 that each hyperedge $e \in E$ is still covered by a vertex $\lambda(v)$ of some vertex $v \in V(T)$.
   - **2nd property of GHDs:** Similar to our proof of Lemma 17, we need to prove that the modified-unique-c-gc inactivation does not locally disconnect any $p$, $s$, $u$, $c$, and $gc$. In addition, now we also need to prove that $s$ does not share any attribute with any other vertex in the tree, that it does not already share with $p$, $u$, $c$, or $gc$. This is because the values of $\chi(s)$ are now assigned through the $X$ values on the edges, and not the $\chi$ values of $p$, $u$, $c$, and $gc$.

     Similar to our proof of Lemma 17, we again have to consider all possible breaks in connectedness between $p$, $u$, $c$, and $gc$ introduced by the insertion of $s$. We only show the proof for attributes between $p$ and $gc$. Consider any attribute $X \in \chi(p) \cap \chi(gc)$. Then, since the initial $D'$ was a valid GHD, $X$ must have been in $\chi(u)$. Then $\chi(s)$ also includes $X$ because $\chi(s)$ by construction contains $X(p,u)$, which by assumption (4-c) contains $\chi(p) \cap \chi(u)$.

     Now assume for the purpose of contradiction that $\chi(s)$ contains an attribute $A$ with a vertex $z$, and $A$ is not in $\chi(p)$, $\chi(u)$, $\chi(c)$, and $\chi(gc)$. Since $\chi(s) = X(p,u) \cup X(u,c) \cup X(c,gc)$, and therefore by assumption (4-d) $\chi(s) \subseteq \chi(u) \cup \chi(c) \cup \chi(gc)$, contradicting the assumption that $A$ is not in $\chi(u)$, $\chi(c)$, or $\chi(gc)$.
   - **3rd property of GHDs:** We only need to prove that $\chi(s) \subseteq \cup \lambda(s)$. $\chi(s) = X(p,u) \cup X(u,c) \cup X(c,gc)$, each of these three components are by assumption (4-b) covered (respectively) by the $\Lambda(u,c)$, $\Lambda(u,c)$, and $\Lambda(c,gc)$, which comprise $\lambda(s)$.
   - **Width of the modified GHD:** By assumption (4-a), the sizes of $\Lambda(u,c)$, $\Lambda(u,c)$, and $\Lambda(c,gc)$ are all at most w. Therefore $|\lambda(s)|$ is at most 3w, showing that the width of the GHD is still at most 3w, completing our proof.

**HD's Descendant Property:** Leaf inactivation does not modify the $\chi$ and $\lambda$ assignments, so trivially maintains the descendant property. For modified-unique-c-gc inactivation, we need to prove that the property holds for $p$, $s$, $u$ and $c$:

- $u$ and $c$: The property holds as $u$ and $c$'s $\lambda$ assignments remain the same and $T_u$ and $T_c$ can only get smaller.
- $p$: The only new vertex in $T_p$ is $s$ and $\chi(s) = X(p,u) \cup X(u,c) \cup X(c,gc)$. Recall that we proved in the proof of the 3rd property of GHDs above that $X(p,u) \cup X(u,c) \cup X(c,gc) \subset \chi(u) \cup \chi(c) \cup \chi(gc)$, implying that any attribute in $s$ were already part of $T_p$.
- $s$: For $s$ we first need to prove a lemma about the relations in $\Lambda(u,v)$, which we call the ***edge descendant property***.

  ▶ **Lemma 32.** *Assume the following property holds in the initial GHD $D'$: $\forall e \in \Lambda(u,v)$, $(attr(e) \cap T_v) \subseteq X(u,v)$, then this property is maintained through a series of leaf and modified-unique-c-gc inactivations.*

  **Proof.** We only need to consider modified-unique-c-gc inactivation. We need to consider that the property holds for $(p,s)$ and $(s,gc)$. We only provide the proof for $(p,s)$.

Consider any edge $e \in \Lambda(p,s) = \Lambda(p,u)$. Notice that $T_s = T_u \cup \chi(s)$. Recall that we proved above that $\chi(s) \subseteq \chi(u) \cup \chi(c) \cup \chi(gc)$. Therefore $T_s = T_u$. Since by assumption $attr(e) \cap T_u \subseteq X(p,u)$, and $X(p,s) = X(p,u)$, $attr(e) \cap T_s \subseteq X(p,s) = X(p,u)$. ◀

Now, consider any $e \in \lambda(s)$. $e$ can come from $\lambda(p,u)$, $\lambda(u,c)$, or $\lambda(c,gc)$. We only prove the case when $e \in \lambda(p,u)$. Then we proved in Lemma 32 that $e \cup T_s = T_u \subseteq X(p,u) \subseteq \chi(s)$, completing the proof.

◀

## G.4 Improving Akatov's and Bodlaender's Results

We can also improve Akatov's and Bodlaender's results if we modify Log-GTA and use similar notions of intersection width that we used for GHDs.

### G.4.1 Improving Bodlaender's Result

Let the **tree intersection width** of a TD be the maximum size of the number of attributes shared between two adjacent vertices in the TD. Further, let the tree intersection width of a hypergraph $H$ with treewidth tw be the minimum tree intersection width of any of its tw-treewidth TDs. For example, the GHD we showed for $TC_n$ from Figure 1c is also a TD, and has treewidth 2 and tree intersection width of 1.

Let $D$ be a TD with treewidth tw and tree intersection width of tiw. Notice that by definition tiw is at most $\text{tw} + 1$. Then using Log-GTA, we can actually improve Bodlaender's result from $3\text{tw} + 2$ to $\max(\text{tw}, 3\text{tiw} - 1)$. We note that we do not modify Log-GTA. We only modify our analysis using the notion of tree intersection width to get this strictly better result.

We start with a TD of $H$ with tree intersection width tiw (one must exist by definition). Observe that the number of shared attributes between any two vertices during Log-GTA is always at most tiw. To see that this property is maintained throughout Log-GTA, we only need to prove that $|\chi(p) \cap \chi(s)| \leq \text{tiw}$ and $|\chi(s) \cap \chi(gc)| \leq \text{tiw}$. We showed in our proof of Lemma 17 (property 4 of unique-c-gc inactivation case), that $\chi(p) \cap \chi(s) = \chi(p) \cap \chi(u)$, which by assumption is at most tiw. The situation is similar for $\chi(s) \cap \chi(gc)$. Therefore the $|\chi(s)|$, which is the union of three intersections, is at most 3tiw. At the end of the transformation, all vertices that have never been an $s$ vertex during the transformation have at most $\text{tw} + 1$ attributes, and those that have been $s$ vertices have 3tiw attributes. Therefore the treewidth of the final $D'$ is $\max(\text{tw}, 3\text{tiw} - 1)$. For example, we can check that Log-GTA's transformation of $TC_{15}$, shown in Figure 8, yields a TD with treewidth of $\max(2, 3 - 1) = 2$, instead of 8.

### G.4.2 Improving Akatov's Result

Similar to our definition of intersection width, let the **hypertree intersection width** of a HD $D = (T, \chi, \lambda)$ be defined as follows: For any adjacent vertices $u, v \in V(T)$, let $\text{iw}(u,v)$ denote the size of the smallest set $S \subseteq E(H)$ such that $\chi(u) \cap \chi(v) \subseteq \bigcup_{s \in S} s$. In other words, $\text{iw}(u,v)$ is the size of the smallest set of relations whose attributes cover the common attributes between $u$ and $v$. The hypertree intersection width hiw of an HD is the maximum $\text{iw}(u,v)$ over all adjacent $u, v \in V(T)$. For example, the GHD we showed for $TC_{15}$ from Figure 1c is also an HD, and has width 2 and hypertree intersection width 1.

To improve Akatov's result, we need to modify the Log-GTA algorithm. Let $D$ be an $HD$ with width w and hypertree intersection width of hiw Now consider making the following two modifications to Log-GTA: (1) When extending the initial HD $D$ to $D'$, we assign the common covers of the edges to be any covering subset of size hiw; and (2) During unique-c-gc inactivation operation, instead of assigning $\chi(s)$ to be $(\chi(p) \cap \chi(u)) \cup (\chi(u) \cap \chi(c)) \cup (\chi(c) \cap \chi(gc))$, we assign $\chi(s) = attr(\lambda(p,u)) \cup attr(\lambda(u,c)) \cup attr(\lambda(c,gc))$. We omit the proof, but it can be shown that this modified Log-GTA returns a $O(\log(n))$-depth $D'$ with width $\max(\mathsf{w}, 3\mathsf{hiw})$, improving Akatov's result from $3\mathsf{w}$ to $\max(\mathsf{w}, 3\mathsf{hiw})$. For example, in the case of $TC_{15}$, shown in Figure 8, the returned HD would have a width of 3, instead of Akatov's result of 6.